\documentstyle[preprint,aps,epsfig]{revtex}
\newcommand{\lsim}{\mathrel{\rlap{\lower4pt\hbox{\hskip0pt$\sim$}}
\raise1pt\hbox{$<$}}}
\tighten
\begin{document}
%
\preprint{Preprint Numbers: \parbox[t]{45mm}{ANL-PHY-9217-TH-98\\
                                             nucl-th/9812063}}
 \title{Survey of heavy-meson observables}
\author{M. A. Ivanov\footnotemark[1], Yu. L. Kalinovsky\footnotemark[2] and
 C. D. Roberts\footnotemark[3]\vspace*{0.2\baselineskip}}

 \address{\footnotemark[1]Bogoliubov Laboratory of Theoretical Physics,
 \\Joint Institute for Nuclear Research, 141980 Dubna,
 Russia\vspace*{0.2\baselineskip}\\  
 \footnotemark[2]Laboratory of Computing Techniques and Automation, \\Joint
 Institute for Nuclear Research, 141980 Dubna,
 Russia\vspace*{0.2\baselineskip}\\  
 \footnotemark[3]Physics Division, Bldg. 203, Argonne National Laboratory,
 Argonne IL 60439-4843\vspace*{0.2\baselineskip}}
\date{Pacs Numbers: 13.20.-v, 13.20.Fc, 13.20.He, 24.85.+p}
\maketitle
\begin{abstract}
We employ a Dyson-Schwinger equation model to effect a unified and uniformly
accurate description of light- and heavy-meson observables, which we
characterise by heavy-meson leptonic decays, semileptonic heavy-to-heavy and
heavy-to-light transitions - $B \to D^\ast$, $D$, $\rho$, $\pi$; $D \to
K^\ast$, $K$, $\pi$, radiative and strong decays - $B_{(s)}^\ast \to
B_{(s)}\gamma$; $D_{(s)}^*\to D_{(s)}\gamma$, $D \pi$, and the rare $B\to
K^\ast \gamma$ flavour-changing neutral-current process.  We elucidate the
heavy-quark limit of these processes and, using a model-independent mass
formula valid for all nonsinglet pseudoscalar mesons, demonstrate that their
mass rises linearly with the mass of their heaviest constituent.  In our
numerical calculations we eschew a heavy-quark expansion and rely instead on
the observation that the dressed $c,b$-quark mass functions are well
approximated by a constant, interpreted as their constituent-mass: we find
$\hat M_c=1.32\,$GeV and $\hat M_b=4.65\,$GeV.  The calculated heavy-meson
leptonic decay constants and transition form factors are a necessary element
in the experimental determination of CKM matrix elements.  The results also
show that this framework, as employed hitherto, is well able to describe
vector meson polarisation observables.
\end{abstract}
\section{Introduction}
\label{intro}
Mesons are the simplest bound states in QCD, and their non-hadronic
electroweak interactions provide an important tool for exploring their
structure and elucidating the nonperturbative, long-distance behaviour of the
strong interaction.  That elucidation is accomplished most effectively by
applying a single framework to a broad range of observables, and in this the
bound state phenomenology~\cite{peter,cdranu} based on Dyson-Schwinger
equations (DSEs)~\cite{review} has been successful; e.g., with its
simultaneous application to phenomena as diverse as $\pi$-$\pi$
scattering~\cite{cdranom,pipi}, the electromagnetic form factors of light
pseudoscalar mesons~\cite{cdrpion,brtpik,mrpion}, anomalous
pion~\cite{cdranom} and photo-pion~\cite{mrpion,photopi,dubravko} processes,
and the diffractive electroproduction~\cite{pichowsky} and electromagnetic
form factors~\cite{ph98} of vector mesons.  Herein we extend this application
and implement a simplification valid for heavy quarks, so obtaining in
addition a description of heavy-meson observables.  We illustrate that by
reporting the simultaneous calculation of a range of light-meson observables
and heavy-meson leptonic decays, semileptonic heavy-to-heavy and
heavy-to-light transitions - $B \to D^*$, $D$, $\rho$, $\pi$; $D \to K^\ast$,
$K$, $\pi$, radiative and strong decays - $B_{(s)}^\ast \to B_{(s)}\gamma$;
$D_{(s)}^*\to D_{(s)}\gamma$, $D \pi$, and the rare $B\to K^\ast \gamma$
flavour-changing neutral-current process.  This is an extensive but not
exhaustive range of applications.

To introduce the heavy-quark simplification we observe that mesons, whether
heavy or light, are bound states of a dressed-quark and -antiquark, where the
dressing is described by the quark Dyson-Schwinger equation
(DSE)~\cite{review}:\footnote{We use a Euclidean formulation with
$\{\gamma_\mu,\gamma_\nu\}=2\delta_{\mu\nu}$, $\gamma_\mu^\dagger =
\gamma_\mu$ and $p\cdot q=\sum_{i=1}^4 p_i q_i$.  A vector, $k_\mu$, is
timelike if $k^2<0$.}
\begin{eqnarray}
\label{genS}
S_f(p)^{-1} & := & 
i \gamma\cdot p \,A_f(p^2) + B_f(p^2)  =
A_f(p^2) \left( i \gamma\cdot p + M_f(p^2) \right)\\
\label{gendse} & = & Z_2 (i\gamma\cdot p + m_f^{\rm bm})
+\, Z_1\, \int^\Lambda_q \,
g^2 D_{\mu\nu}(p-q) \frac{\lambda^a}{2}\gamma_\mu S_f(q)
\Gamma^{fa}_\nu(q,p) \,.
\end{eqnarray}
Here $f\,(=u,d,s,c,b)$ is a flavour label, $D_{\mu\nu}(k)$ is the
dressed-gluon propagator, $\Gamma^{fa}_\nu(q,p)$ is the dressed-quark-gluon
vertex, $m_f^{\rm bm}$ is the $\Lambda$-dependent current-quark bare mass and
$\int^\Lambda_q := \int^\Lambda d^4 q/(2\pi)^4$ represents mnemonically a
{\em translationally-invariant} regularisation of the integral, with
$\Lambda$ the regularisation mass-scale.  The renormalisation constants for
the quark-gluon-vertex and quark wave function, $Z_1(\zeta^2,\Lambda^2)$ and
$Z_2(\zeta^2,\Lambda^2)$, depend on the renormalisation point, $\zeta$, and
the regularisation mass-scale, as does the mass renormalisation constant
$Z_m(\zeta^2,\Lambda^2) := Z_2(\zeta^2,\Lambda^2)^{-1}
Z_4(\zeta^2,\Lambda^2)$.  However, one can choose the renormalisation scheme
such that they are flavour-independent.

This equation has been much studied and the qualitative features of its
solution elucidated.  In QCD the chiral limit is defined by $\hat m = 0$,
where $\hat m$ is the renormalisation-point-independent current-quark mass.
It follows that in this case there is no scalar, mass-like divergence in the
perturbative evaluation of the quark self energy.  Hence, for
$p^2>20\,$GeV$^2$ the solution of Eq.~(\ref{gendse}) for the chiral-limit
quark mass-function is~\cite{mr97}
\begin{equation}
\label{Mchiral}
M_0(p^2) \stackrel{{\rm large}-p^2}{=}\,
\frac{2\pi^2\gamma_m}{3}\,\frac{\left(-\,\langle \bar q q \rangle^0\right)}
           {p^2
        \left(\case{1}{2}\ln\left[p^2/\Lambda_{\rm QCD}^2\right]
        \right)^{1-\gamma_m}}\,,
\end{equation}
where $\gamma_m=12/(33-2 N_f)$ is the gauge-independent mass anomalous
dimension and $\langle \bar q q \rangle^0$ is the
renormalisation-point-independent vacuum quark condensate.  The existence of
dynamical chiral symmetry breaking (DCSB) means that $\langle \bar q q
\rangle^0 \neq 0$, however, its actual value depends on the long-range
behaviour of $D_{\mu\nu}(k)$ and $\Gamma^{0a}_\nu(q,p)$, which is modelled in
contemporary DSE studies.  $\langle \bar q q \rangle^0\approx -
(0.24\,\mbox{GeV})^3$ is consistent with light-meson
observables~\cite{derek}.

In contrast, for $\hat m_f \neq 0$,
\begin{equation}
\label{masanom}
M_f(p^2) \stackrel{{\rm large}-p^2}{=} \frac{\hat m_f}
{\left(\case{1}{2}\ln\left[p^2/\Lambda_{\rm
QCD}^2\right]\right)^{\gamma_m}}\,.
\end{equation}
An obvious qualitative difference is that, relative to Eq.~(\ref{masanom}),
the chiral-limit solution is $1/p^2$-suppressed in the ultraviolet.

\begin{figure}[t]
\centering{\
\epsfig{figure=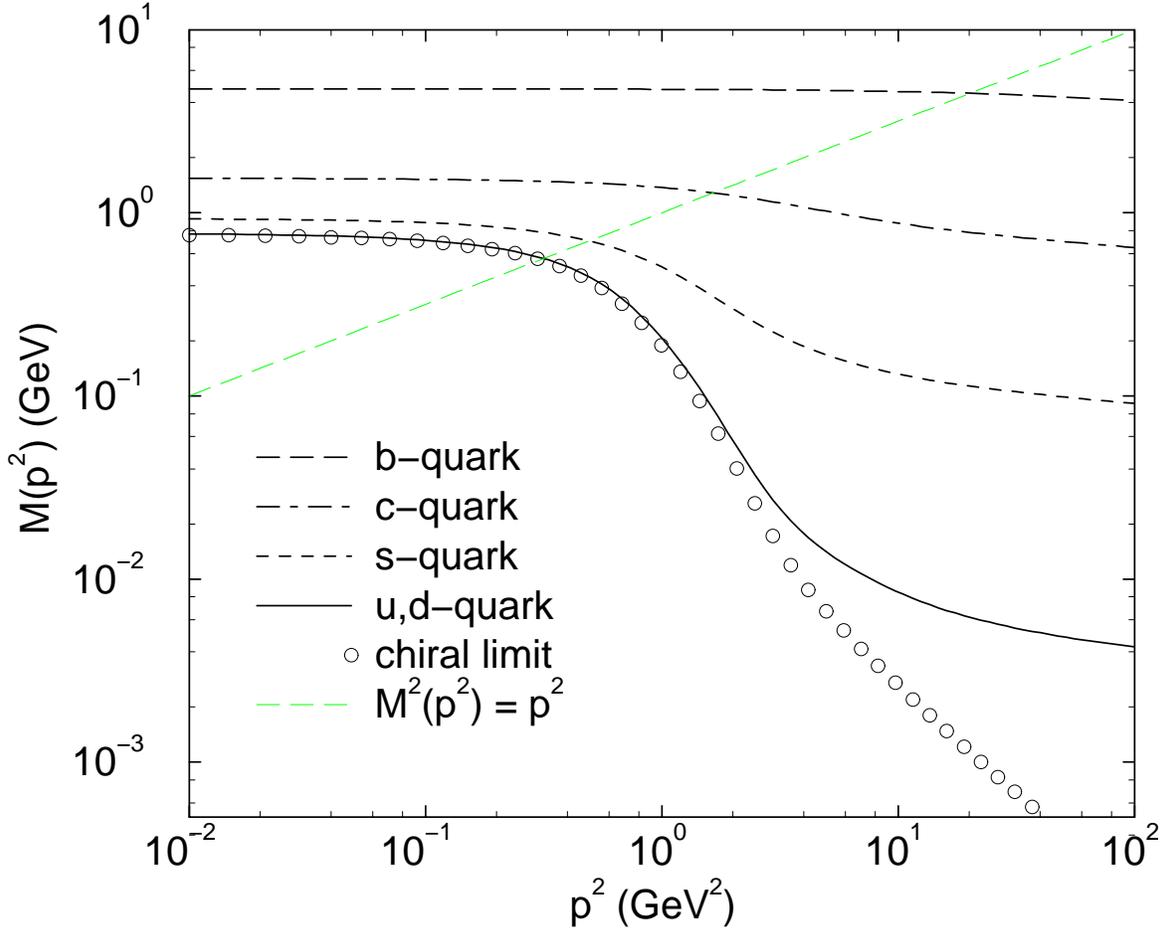,height=12.5cm}}
\caption{Quark mass function obtained as a solution of
Eq.~(\protect\ref{gendse}) using $D_{\mu\nu}(k)$ and $\Gamma^{fa}_\nu(q,p)$
from Ref.~\protect\cite{mr97} and current-quark masses, fixed at $\zeta=
19\,$GeV: $m_{u,d}^\zeta = 3.7\,$MeV, $m_s^\zeta = 82\,$MeV,
$m_c^\zeta=0.58\,$GeV and $m_b^\zeta=3.8\,$GeV.
The indicated solutions of $M^2(p^2)=p^2$ define the Euclidean
constituent-quark mass, $M^E_f$, which takes the values: $M^E_u=0.56\,$GeV,
$M^E_s=0.70\,$GeV, $M^E_c= 1.3\,$GeV, $M^E_b= 4.6\,$GeV.
\label{mp2fig}}
\end{figure}

There is some quantitative model-dependence in the momentum-evolution of the
mass-function into the infrared.  However, for any forms of $D_{\mu\nu}(k)$
and $\Gamma^{fa}_\nu(q,p)$ that provide an accurate description of
$f_{\pi,K}$ and $m_{\pi,K}$, one obtains~\cite{mr98} quark mass-functions
with profiles like those illustrated in Fig.~\ref{mp2fig}.  The evolution to
coincidence between the chiral-limit and $u,d$-quark mass functions, apparent
in this figure, makes clear the transition from the perturbative to the
nonperturbative domain.  The chiral limit mass-function is nonzero {\it only}
because of the nonperturbative DCSB mechanism whereas the $u,d$-quark mass
function is purely perturbative at $p^2>20\,$GeV$^2$, where
Eq.~(\ref{masanom}) is accurate.  The DCSB mechanism thus has a significant
effect on the propagation characteristics of $u,d,s$-quarks.

However, as evident in the figure, that is not the case for the $b$-quark.
Its large current-quark mass almost entirely suppresses momentum-dependent
dressing, so that $M_b(p^2)$ is nearly constant on a substantial domain.
This is true to a lesser extent for the $c$-quark.

We employ ${\cal L}_f:=M^E_f/m_f^\zeta$ as a single, quantitative measure of
the importance of the DCSB mechanism; i.e., nonperturbative effects, in
modifying the propagation characteristics of a given quark flavour.  In this
particular illustration it takes the values:
\begin{equation}
\label{Mmratio}
\begin{array}{c|ccccc}
        f   &   u,d  &   s   &  c  &  b  \\ \hline
 {\cal L}_f &  150   &    10      &  2.2 &  1.2 
\end{array}\,.
\end{equation}
These values are representative: for light-quarks ${\cal L}_{q=u,d,s} \sim
10$-$100$, while for heavy-quarks ${\cal L}_{Q=c,b} \sim 1$, and highlight
the existence of a mass-scale characteristic of DCSB: $M_\chi$.  The
propagation characteristics of a flavour with $m_f^\zeta\leq M_\chi$ are
significantly altered by the DCSB mechanism, while for flavours with
$m_f^\zeta\gg M_\chi$ momentum-dependent dressing is almost irrelevant.  It
is apparent and unsurprising that $M_\chi \sim 0.2\,$GeV$\,\sim \Lambda_{\rm
QCD}$.  As a consequence we anticipate that the propagation of $c,b$-quarks
can be described well by replacing their mass-functions with a constant;
i.e., writing
\begin{equation}
\label{dsehq}
S_Q(p) = \frac{1}{i\gamma \cdot p + \hat M_Q}\,,\; Q=c,b\,,
\end{equation}
where $\hat M_Q$ is a constituent-heavy-quark mass
parameter.\footnote{Although not illustrated explicitly, when $M_f(p^2)
\approx\,$const., $A_f(p^2)\approx 1$ in Eq.~(\protect\ref{genS}).
Eq.~(\protect\ref{dsehq}) is an implicit assumption in the formulation of
Bethe-Salpeter equation models of heavy-mesons, such as Ref.~\cite{vary97}.}
We expect that a good description of observable phenomena will require $\hat
M_Q\approx M_Q^E$.

When considering a meson with a heavy-quark constituent one can proceed
further, as in heavy-quark effective theory (HQET)~\cite{neubert94}, allow
the heaviest quark to carry all the heavy-meson momentum: 
\begin{equation}
P_\mu=: m_H v_\mu=: (\hat M_Q + E_H)v_\mu, 
\end{equation}
and write
\begin{equation}
\label{hqf}
S_Q(k+P) = \case{1}{2}\,\frac{1 - i \gamma\cdot v}{k\cdot v - E_H}
+ {\rm O}\left(\frac{|k|}{\hat M_{Q}},
                \frac{E_H}{\hat M_{Q}}\right)\,,
\end{equation}
where $k$ is the momentum of the lighter constituent.  In the calculation of
observables, the meson's Bethe-Salpeter amplitude will limit the range of
$|k|$ so that Eq.~(\ref{hqf}) will only be a good approximation if {\it both}
the momentum-space width of the amplitude, $\omega_H$, and the binding
energy, $E_H$, are significantly less than $\hat M_Q$.

In Ref.~\cite{mishaB} the propagation of $c$- and $b$-quarks was described by
Eq.~(\ref{hqf}), with a goal of exploring the fidelity of that idealisation.
It was found to allow for a uniformly good description of $B_f$-meson
leptonic and semileptonic decays with heavy- and light-pseudoscalar final
states.  In that study, corrected as described below, $\omega_{B_f} \approx
1.3\,$GeV and $E_{B_f} \approx 0.70\,$GeV, both of which are small compared
with $\hat M_b \approx 4.6\,$GeV in Fig.~\ref{mp2fig}, so that the accuracy
of the approximation could be anticipated.  It is reasonable to expect that
$\omega_D\approx \omega_B$ and $E_D\approx E_B$, since they must be the same
in the limit of exact heavy-quark symmetry.  Hence in processes involving the
weak decay of a $c$-quark ($\hat M_c\approx 1.3\,$GeV) where a $D_f$-meson is
the heaviest participant, Eq.~(\ref{hqf}) must be inadequate; an expectation
verified in Ref.~\cite{mishaB}.

The failure of Eq.~(\ref{hqf}) for the $c$-quark complicates or precludes the
development of a unified understanding of $D_f$- and $B_f$-meson observables
using such contemporary theoretical tools as HQET and light cone sum rules
(LCSRs)~\cite{braun}.  However, the constituent-like dressed-heavy-quark
propagator of Eq.~(\ref{dsehq}) can still be used to effect a unified and
accurate simplification of the study of these observables.  Herein, to
demonstrate this, we extend Refs.~\cite{mishaB,mishaPLB} and employ
Eq.~(\ref{dsehq}), with $\hat M_Q$ treated as free parameters, and
parametrisations~\cite{cdrpion,brtpik,mrpion,photopi,pichowsky,ph98} of the
dressed-light-quark propagators and meson Bethe-Salpeter amplitudes in the
calculation of a wide range of observables, determining the parameters in a
$\chi^2$-fit to a subset of them.  It is an efficacious strategy.

Our article is divided into eight sections with a single appendix.  We
discuss heavy- and light-meson leptonic decays in Sec.~\ref{secb}, and their
masses in Sec.~\ref{secc}.  In Sec.~\ref{secd} we introduce the impulse
approximation to the semileptonic decays of heavy-mesons, and describe the
light-quark propagators and meson Bethe-Salpeter amplitudes necessary for
their evaluation.  The impulse approximation to the other processes is
presented in Sec.~\ref{sece}, while in Sec.~\ref{secf} we elucidate the
heavy-quark symmetry limits of all the decays and transitions.  The accuracy
of these heavy-quark symmetry predictions is discussed in conjunction with
the complete presentation of our results in Sec.~\ref{secg} and
Sec.~\ref{sech} contains some concluding remarks.

\section{Leptonic Decays}
\label{secb}
\subsection{Pseudoscalar Mesons}
The leptonic decay of a pseudoscalar meson, $P(p)$, is described by the
matrix element~\cite{mr97}
\begin{eqnarray}
\label{fwk}
f_{P} \,p_\mu := \langle 0| \bar {\cal Q}\, (T^P)^{\rm T}
\gamma_\mu \gamma_5 \,{\cal Q} | P(p) \rangle 
= {\rm tr}\,Z_2 \int^\Lambda_k \,(T^P)^{\rm T}\gamma_5\gamma_\mu\,
        \chi_P(k;p),
\end{eqnarray}
where 
$\chi_P(k;p)= {\cal S}(k+p)\,\Gamma_P(k;p)\, {\cal S}(k)$,
${\cal Q}= {\rm column}(u,d,s,c,b)$, $T^P$ is a flavour matrix identifying
the meson; e.g., $T^{\pi^+}= \case{1}{2}\left(\lambda^1 + i
\lambda^2\right)$, $(T^P)^{\rm T}$ is its transpose, ${\cal S}= {\rm
diag}(S_u,S_d,S_s,S_c,S_b)$, and the trace is over colour, Dirac and flavour
indices.  $\Gamma_P$ is the meson's Bethe-Salpeter amplitude, which is
normalised canonically according to:
\begin{eqnarray}
\nonumber \lefteqn{
2 \,p_\mu  = 
{\rm tr}\int^\Lambda_q\,\bar\Gamma_P(q;-p)\,
        \frac{\partial {\cal S}(q+p)}{\!\!\!\!\!\!\!\!\!\!\!\!\partial p_\mu}\,
        \Gamma_P(q;p) \,{\cal S}(q) }\\
& & +
\int^\Lambda_{q}\int^\Lambda_{k} \,[\bar\chi_P(q;-p)]_{sr} 
\frac{\partial K^{rs}_{tu}(q,k;p)}
{\!\!\!\!\!\!\!\!\!\!\!\!\partial p_\mu}\, 
[\chi_P(k;p)]_{ut}\,,
\label{psnorm}
\end{eqnarray}
where: $\bar \Gamma_P^{\rm T}(k,-p) = C^{-1} \Gamma_P(-k,-p) C$, with
$C=\gamma_2 \gamma_4$, the charge conjugation matrix; $r,s,t,u$ are colour-,
Dirac- and flavour-matrix indices; and $K$ is the quark-antiquark scattering
kernel.  Equation~(\ref{fwk}) is {\it exact} in QCD: the $\Lambda$-dependence
of $Z_2$ ensures that the right-hand-side (r.h.s.) is finite as $\Lambda \to
\infty$, and its $\zeta$- and gauge-dependence is just that necessary to
compensate that of $\chi_P(k;p)$.

\begin{figure}[t]
\centering{\
\epsfig{figure=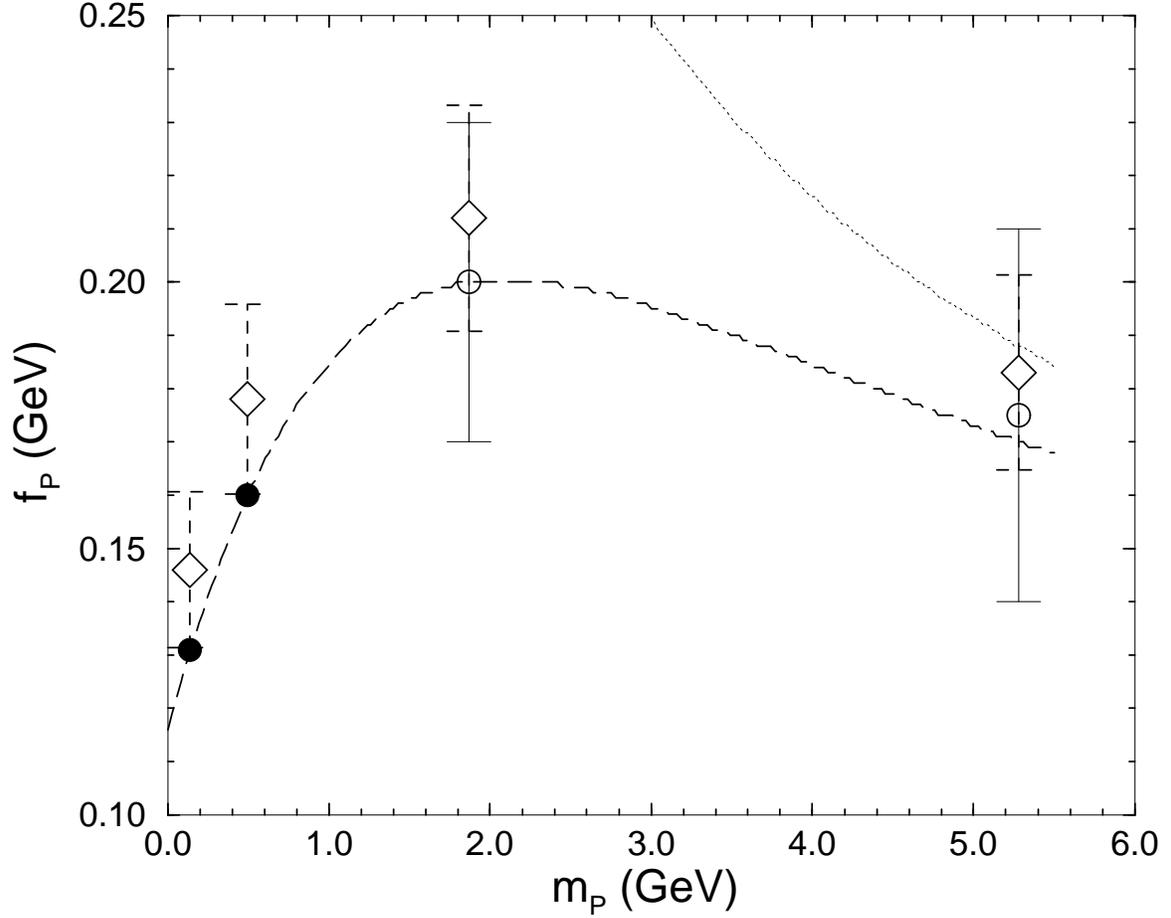,height=12.5cm}}
\caption{Experimental values of $f_{\pi,K}$, filled circles; lattice
estimates of $f_{D,B}$ in Eq.~(\protect\ref{latticefwk}), open circles; our
calculated values of $f_{\pi,K,D,B}$, diamonds, see Sec.~\protect\ref{secg}.
(We estimate a theoretical error of 10\%.)  The dashed line is a fit to the
experimental values and lattice estimates: $f_P^2= (0.013 + 0.028 \,m_P )/(1+
0.055\,m_P + 0.15 \,m_P^2)$, which exhibits the large-$m_P$ limit of
Eq.~(\protect\ref{fwkasymp}), and the dotted line is the large-$m_P$ limit of
this fit.\label{figfwk}}
\end{figure}

The leptonic decay constants of light pseudoscalar mesons, $\pi$, $K$, are
known~\cite{pdg98}: $f_\pi = 0.131\,$GeV, $f_K=0.160\,$GeV.  The increase
with increasing current-quark mass is easily reproduced in DSE
studies~\cite{mr97} and continues until at least $3\,\hat
m_s$~\cite{marisPC}, at which point the renormalisation-group-improved and
confining ladder-like truncation of $K$ used in those studies becomes
inadequate, and that model no longer allows the $m_P$-dependence of $f_P$ to
be tracked directly.  However, we note from Eq.~(\ref{psnorm}) that
\begin{equation}
\label{calGP}
{\cal G}_P(k;p) := \frac{1}{\surd m_P}\,
\Gamma_P(k;p) < \infty\,, m_P \to \infty\,;
\end{equation}
i.e., that ${\cal G}_P(k;p)$ is mass-independent in the heavy-quark symmetry
limit, and hence it follows~\cite{mishaPLB} from the general form of the
meson Bethe-Salpeter amplitude and Eqs.~(\ref{hqf}) -~(\ref{psnorm}) that for
large pseudoscalar meson masses
\begin{equation}
\label{fwkasymp}
f_P \propto 1/\surd{m_P}\,.
\end{equation}
In this model-independent result we recover a well-known general consequence
of heavy-quark symmetry~\cite{neubert94}.  However, the value of the
current-quark mass at which it becomes evident is unknown in spite of the
many studies that report values of $f_D$ and $f_B$, some of which are
tabulated in Ref.~\cite{richman}, and the results of lattice simulations, a
summary~\cite{flynn} of which reports
\begin{equation}
\label{latticefwk}
f_D = 200 \pm 30\,{\rm MeV}\,,\; f_B= 170 \pm 35\,{\rm MeV}\,.
\end{equation}
In Fig.~\ref{figfwk} we illustrate the behaviour of $f_P$ we {\it anticipate}
based on these observations.  It suggests that D-mesons lie outside the
domain on which Eq.~(\protect\ref{fwkasymp}) is manifest.

\subsection{Vector Mesons}
The leptonic decay of a vector meson, $V_\lambda(p)$, is described by the
matrix element ($\epsilon_\mu^\lambda(p)$ is the polarisation vector:
$\epsilon^\lambda(p)\cdot p=0$)
\begin{eqnarray}
\nonumber
\lefteqn{f_V\,M_V\,\epsilon^\lambda_\mu(p):= 
\langle 0| \bar {\cal Q}\, (T^P)^{\rm T} \gamma_\mu  \,
        {\cal Q} | V_\lambda(p) \rangle\,,}\\
&& \label{fwkV}
\Rightarrow \, f_V\, M_V  = 
\case{1}{3} \,{\rm tr}\,Z_2\int^\Lambda_k \, (T^P)^{\rm T}\gamma_\mu\,
\chi^V_{\mu}(k;p)\,,
\end{eqnarray}
where $\chi^V_{\mu}(k;p)= {\cal S}(k+p)\,\Gamma^V_{\mu}(k;p)\, {\cal S}(k)$
with $\Gamma^V_{\mu}(k;p)$ the vector meson Bethe-Salpeter amplitude, which
is transverse:
\begin{equation}
p_\mu\,\Gamma^V_{\mu}(k;p)=0\,,\; p^2=-M_V^2\,,
\end{equation}
and normalised according to 
\begin{eqnarray}
\nonumber \lefteqn{
2 \,p_\mu  = 
\case{1}{3}\, {\rm tr}\int^\Lambda_q\,\bar\Gamma^V_\nu(q;-p)\,
        \frac{\partial {\cal S}(q+p)}{\!\!\!\!\!\!\!\!\!\!\!\!\partial p_\mu}\,
        \Gamma^V_\nu(q;p) \,{\cal S}(q) }\\
& & +
\int^\Lambda_{q}\int^\Lambda_{k} \,[\bar\chi^V_\nu(q;-p)]_{sr} 
\frac{\partial K^{rs}_{tu}(q,k;p)}
{\!\!\!\!\!\!\!\!\!\!\!\!\partial p_\mu}\, 
[\chi^V_\nu(k;p)]_{ut}\,,
\label{vnorm}
\end{eqnarray}
an analogue of Eq.~(\ref{psnorm}).  The obvious analogue of Eq.~(\ref{calGP})
is true.
        
Such decays are difficult to observe directly but it is possible to estimate
$f_\rho$ and hence identify the natural scale of $f_V$.  In the
isospin-symmetric limit the $\rho^0 \to e^+ e^- $ decay constant, $g_\rho$,
is obtained from the matrix element
\begin{equation}
\frac{M_{\rho}^2}{g_\rho}\,\epsilon^\lambda_\mu(p)
:= \langle 0| \bar u \gamma_\mu  u | \rho^0_\lambda(p) \rangle
 =  \case{1}{\surd 2} \, 
\langle 0| \bar u \gamma_\mu  d | \rho^-_\lambda(p) \rangle
\end{equation}
so that
\begin{equation}
f_\rho = \surd 2 \, \frac{M_\rho}{g_\rho}\,.
\end{equation}
The experimentally measured width~\cite{pdg98} $\Gamma_{\rho^0\to e^+ e^-} =
6.77 \pm 0.32\,$keV yields $g_\rho = 5.03 \pm 0.12$ and hence
\begin{equation}
f_\rho = 216 \pm 5 \, {\rm MeV}\,.
\end{equation}

For the vector-mesons it follows from the general form of the Bethe-Salpeter
amplitude, and Eqs.~(\ref{hqf}), (\ref{fwkV}) and (\ref{vnorm}) that for
large vector-meson masses
\begin{equation}
\label{fwkVasymp}
f_V \propto 1/\surd{M_V}\,,
\end{equation}
which again is a model-independent and general consequence of heavy-quark
symmetry.  In fact, as we illustrate below,
\begin{equation}
\label{fpvasymp}
f_P = f_V \propto 1/\surd \hat m_Q\,, \; \hat m_Q \to \infty\,;
\end{equation}
i.e., observables are spin-independent in the heavy-quark symmetry
limit~\cite{neubert94}.

\section{Pseudoscalar Meson Masses}
\label{secc}
Flavour nonsinglet pseudoscalar meson masses satisfy~\cite{mr97}
\begin{eqnarray}
\label{massP}
f_P \, m_P^2 & = & {\cal M}_P^\zeta\,r^\zeta_P\,,\;
{\cal M}_P^\zeta:= {\rm tr}_{\rm flavour} \left[
        M^\zeta\,\left\{T^P,(T^P)^{\rm T}\right\}\right]\,,
\end{eqnarray}
with $M^\zeta= {\rm diag}( m_u^\zeta,m_d^\zeta, m_s^\zeta,m_c^\zeta,
m_b^\zeta)$ and
\begin{eqnarray}
\label{rP}
i\,r_P^\zeta & = &  {\rm tr}\,Z_4\int^\Lambda_k\, 
        \left( T^P \right)^{\rm T}\gamma_5\,\chi_P(k;p)\,.
\end{eqnarray}
The renormalisation constant, $Z_4$, ensures that $r_P^\zeta$ is finite as
$\Lambda \to \infty$, and its $\zeta$- and gauge-dependence is just that
necessary to ensure that the product on the r.h.s.~of Eq.~(\ref{massP}) is
gauge invariant and renormalisation point independent.

In the limit of small current-quark masses one obtains~\cite{mr97} what
is commonly called the Gell-Mann--Oakes--Renner relation as a corollary of
Eq.~(\ref{massP}); i.e., $m_P^2 \propto \hat m_f$, $\hat m_f \to 0$.
However, it also has an important corollary in the heavy-quark symmetry
limit.  Using Eqs.~(\ref{hqf}) and (\ref{calGP})
\begin{equation}
r_P \propto \surd m_P\,, \;m_P \to \infty\,,
\end{equation}
from which Eqs.~(\ref{fwkasymp}) and (\ref{massP}) yield
\begin{equation}
m_P \propto \hat m_Q\,,\; \hat m_Q \to \infty\,.
\end{equation}
Thus the quadratic trajectory, valid when the current-quark mass of the
constituents is small, evolves into a linear trajectory when this mass
becomes large.  In all phenomenologically efficacious DSE models the linear
trajectory is manifest at twice the $s$-quark current-mass~\cite{mr98} so
that $m_K$ lies on the extrapolation of the straight line joining $m_D$ and
$m_B$ in the $(\hat m_f,m_P)$-plane~\cite{cdranu}.

\section{Semileptonic Transition Form Factors}
\label{secd}
\subsection{Pseudoscalar meson in the final state}
The pseudoscalar$\,\to\,$pseudoscalar transition: $P_1(p_1) \to
P_2(p_2)\,\ell\,\nu\,,$ where $P_1$ represents either a $B$ or $D$ and $P_2$
can be a $D$, $K$ or $\pi$, is described by the invariant amplitude
\begin{equation}
A(P_1 \to P_2\,\ell\,\nu)
= \frac{G_F}{\surd 2}\, V_{f^\prime f}\,
\bar \ell \,\gamma_\mu(1 - \gamma_5)\nu\,
M_\mu^{P_1 P_2}(p_1,p_2)\,,
\end{equation} 
where $G_F=1.166 \times 10^{-5}\,$GeV$^{-2}$, $V_{f^\prime f}$ is the relevant
element of the Cabibbo-Kobayashi-Maskawa (CKM) matrix, and the hadronic
current is
\begin{eqnarray}
\label{fpfm}
M_\mu^{P_{1} P_{2}}(p_1,p_2)  & := & 
\langle P_{2}(p_2)| \bar f^\prime \gamma_\mu f | P_{1}(p_1)\rangle
 =  f_+(t) (p_1 + p_2)_\mu + f_-(t) q_\mu\,,
\end{eqnarray}
with $t:= -q^2 = -(p_1-p_2)^2$.  The transition form factors, $f_\pm(t)$,
contain all the information about strong-interaction effects in these
processes, and their accurate calculation is essential for a reliable
determination of the CKM matrix elements from a measurement of the decay
width ($t_\pm := (m_{P_1}\pm m_{P_2})^2$):
\begin{equation}
\label{branching}
\Gamma(P_{1} \to P_{2}\ell\nu)= 
\frac{G_F^2}{192 \pi^3}\,|V_{f^\prime f}|^2\,\frac{1}{m_{P_1}^3}\,
\int_0^{t_-}\,dt\,|f_+(t)|^2\,
\left[(t_+-t) (t_- - t)\right]^{3/2}\!\!.
\end{equation}

\subsection{Vector meson in the final state}
The pseudoscalar$\,\to\,$vector transition: $P(p_1) \to
V_\lambda(p_2)\,\ell\,\nu\,,$ with $P$ either a $B$ or $D$ and $V_\lambda$ a
$D^\ast$, $K^\ast$ or $\rho$, is described by the invariant amplitude
\begin{equation}
A(P \to V_\lambda\,\ell\,\nu) 
= \frac{G_F}{\surd 2}\, V_{f^\prime f}\,
\bar \ell \,i\gamma_\mu(1 - \gamma_5)\nu\,\epsilon_\nu^\lambda(p_2)\,
M_{\mu\nu}^{P V_\lambda}(p_1,p_2)\,,
\end{equation}
where the hadronic tensor involves four scalar functions 
\begin{eqnarray}
\nonumber
\lefteqn{\epsilon_\nu^\lambda(p_2)\,M_{\mu\nu}^{P V_\lambda}(p_1,p_2) =
}\\
&&\nonumber 
\epsilon_\mu^\lambda\,(m_P + M_V)\,A_1(t)
+ (p_1+p_2)_\mu\,\epsilon^\lambda \cdot q\,\frac{A_2(t)}{m_P + M_V}\\
&&
+\,  q_\mu\,\epsilon^\lambda \cdot q\,\frac{A_3(t)}{m_P + M_V}\,
+\, \varepsilon_{\mu\nu\alpha\beta}\, 
        \epsilon_\nu^\lambda \,p_{1\alpha}\, p_{2\beta} \,
        \frac{2 V(t)}{m_P + M_V}\,.
\label{vsl}
\end{eqnarray}
The contribution of $A_3$ can be neglected unless $\ell = \tau$.

Introducing three helicity amplitudes
\begin{eqnarray}
\lefteqn{H_\pm = (m_P + M_V)\, A_1(t)\, \mp \, 
\frac{\lambda^{\frac{1}{2}}(m_P^2,M_V^2,t)}{m_P + M_V}\, V(t)\,,}\\
\nonumber \lefteqn{H_0  = } \\
&&  \!\!\!\!\frac{1}{2\,M_V\,\surd t}\left(
        \left[ m_P^2 - M_V^2 -t\right] \, \left[ m_P + M_V\right] A_1(t)
        - \frac{\lambda(m_P^2,M_V^2,t)}{m_P + M_V}\,A_2(t)\right)\!,
\end{eqnarray}
where $\lambda(m_P^2,M_V^2,t)= [t_+ -t]\, [t_- -t]$, $t_\pm = (m_P\pm
M_V)^2$, the transition rates can be expressed as
\begin{eqnarray}
\frac{d\Gamma_{\pm,0}}{dt} & = & 
\frac{G_F^2}{192 \pi^3 m_{P_1}^3}\,|V_{f^\prime f}|^2\,t\,
\lambda^{\frac{1}{2}}(m_P^2,M_V^2,t)\,|H_{\pm,0}(t)|^2\,,
\end{eqnarray}
and the transverse and longitudinal rates and widths are
\begin{eqnarray}
\frac{d\Gamma_{T}}{dt} & = & 
        \frac{d\Gamma_{+}}{dt} + \frac{d\Gamma_{-}}{dt}\,,\;
\Gamma_T= \int_0^{t_-}\,dt\,\frac{d\Gamma_T}{dt}\,,
\\
\frac{d\Gamma_{L}}{dt} & = & \frac{d\Gamma_{0}}{dt}\,,\;
\Gamma_L= \int_0^{t_-}\,dt\,\frac{d\Gamma_L}{dt}\,,
\end{eqnarray}
with the total width: $\Gamma= \Gamma_T + \Gamma_L$.  The polarisation ratio
and forward-backward asymmetry are
\begin{eqnarray}
\alpha = 2 \,\frac{\Gamma_L}{\Gamma_T} - 1\,, &\;&
A_{\rm FB} = \frac{3}{4}\,\frac{\Gamma_- - \Gamma_+}{\Gamma}\,.
\end{eqnarray}

\subsection{Impulse Approximation}
We employ the impulse approximation in calculating the hadronic contribution
to these invariant amplitudes:
\begin{eqnarray}
\nonumber \lefteqn{{\cal H}^{PX}_\mu(p_1,p_2)  = }\\
&& 2 N_c\,{\rm tr}_D\,\int_k^\Lambda\,
\bar\Gamma_X(k;-p_2)\,S_{q}(k_2) \,
i{\cal O}_\mu^{qQ}(k_2,k_1)\,
S_Q(k_1)\,\Gamma_P(k;p_1)\,S_{q^\prime}(k),
\label{ia}
\end{eqnarray}
where the flavour structure has been made explicit, $k_{1,2} = k+p_{1,2}$
and:
\begin{eqnarray}
{\cal H}^{P_1 X=P_2}_\mu(p_1,p_2) &=& M_\mu^{P_1 P_2}(p_1,p_2)\,,\\
{\cal H}^{P X=V^\lambda}_\mu(p_1,p_2) & = &
\epsilon_\nu^\lambda(p_2)\,M_{\mu\nu}^{P V_\lambda}(p_1,p_2)\,;
\end{eqnarray}
$\Gamma_{X=V^\lambda}(k;p)= \epsilon^\lambda(p)\cdot \Gamma^V(k;p)$; and
${\cal O}_\mu^{qQ}(k_2,k_1)$ is the dressed-quark-W-boson vertex, which in
weak decays of heavy-quarks is well approximated~\cite{mishaB,mishaPLB} by
\begin{equation}
{\cal O}_\mu^{qQ}(k_2,k_1) = \gamma_\mu\,(1 - \gamma_5)\,.
\end{equation}

The impulse approximation has been used widely and efficaciously in
phenomenological DSE studies; e.g.,
Refs.~\cite{cdrpion,brtpik,mrpion,photopi,dubravko,pichowsky,ph98,mishaB,mishaPLB}.
It is self-consistent only if the quark-antiquark scattering kernel is
independent of the total momentum.  Corrections can be incorporated
systematically~\cite{brs96} and their effect in processes such as those
considered herein has been estimated~\cite{mitchell,piloop}: they vanish with
increasing spacelike momentum transfer and contribute $\lsim 15$\% even at
the extreme kinematic limit, $t=t_-$.

\subsection{Quark Propagators}
To evaluate ${\cal H}^{P_1 X}_\mu(p_1,p_2)$ a specific form for the
dressed-quark propagators is required.  As argued in Sec.~\ref{intro},
Eq.(\ref{dsehq}) provides a good approximation for the heavier quarks,
$Q=c,b$, and we use that herein with $\hat M_Q$ treated as free parameters.
For the light-quark propagators:
\begin{equation}
\label{defS}
S_f(p) = -i\gamma\cdot p\,\sigma^f_V(p^2) + \sigma^f_S(p^2)
        = \frac{1}{i\gamma\cdot p\,A_f(p^2) + B_f(p^2)}\,,
\end{equation}
$f=u,s$ (isospin symmetry is assumed), we use the algebraic forms introduced
in Ref.~\cite{cdrpion}, which efficiently characterise the essential and
robust elements of the solution of Eq.~(\ref{gendse}) and have been used
efficaciously in
Refs.~\cite{cdrpion,brtpik,mrpion,photopi,pichowsky,ph98,mishaB,mishaPLB}:
\begin{eqnarray}
\label{SSM}
\bar\sigma^f_S(x)  & =  & 
        2 \bar m_f {\cal F}(2 (x + \bar m_f^2))
        + {\cal F}(b_1 x) {\cal F}(b_3 x) 
        \left( b^f_0 + b^f_2 {\cal F}(\epsilon x)\right)\,,\\
\label{SVM}
\bar\sigma^f_V(x) & = & \frac{2 (x+\bar m_f^2) -1 
                + e^{-2 (x+\bar m_f^2)}}{2 (x+\bar m_f^2)^2}\,,
\end{eqnarray}
${\cal F}(y)= (1-{\rm e}^{-y})/y$, $x=p^2/\lambda^2$; $\bar m_f$ =
$m_f/\lambda$; and
\begin{eqnarray}
\bar\sigma_S^f(x)  =  \lambda\,\sigma_S^f(p^2)\,,&\;&
\bar\sigma_V^f(x)  =  \lambda^2\,\sigma_V^f(p^2)\,,
\end{eqnarray}
with $\lambda$ a mass scale.  This algebraic form combines the effects of
confinement\footnote{The representation of $S(p)$ as an entire function is
motivated by the algebraic solutions of Eq.~(\protect\ref{gendse}) in
Refs.~\cite{munczekburden} and the concomitant absence of a Lehmann
representation is a sufficient condition for
confinement~\protect\cite{cdranu,review,brs96}.} and DCSB with free-particle
behaviour at large, spacelike $p^2$.\footnote{At large-$p^2$: $\sigma_V(p^2)
\sim 1/p^2$ and $\sigma_S(p^2)\sim m/p^2$.  Therefore the parametrisation
does not incorporate the additional $\ln p^2$-suppression characteristic of
QCD.  It is a useful simplification, which introduces model artefacts that
are easily identified and accounted for.  $\varepsilon=10^{-4}$ is introduced
only to decouple the large- and intermediate-$p^2$ domains.}

The chiral limit vacuum quark condensate is~\cite{mr97} 
\begin{eqnarray}
-\langle \bar q q \rangle^0_\zeta & := & 
\lim_{\Lambda^2\to\infty}\,N_c\,{\rm tr}_{D}\,Z_4(\zeta^2,\Lambda^2)\,
\int_k^\Lambda\, S_0(k),
\label{condensate}
\end{eqnarray}
where at one-loop order
$ Z_4(\zeta^2,\Lambda^2) = [\alpha(\Lambda^2)/\alpha(\zeta^2)]^{\gamma_m ( 1
+ \xi/3)}$,
with $\xi$ the co\-va\-riant-gauge parameter ($\xi=0$ specifies Landau
gauge).  The $\xi$-dependence of $ Z_4(\zeta^2,\Lambda^2) $ is just that
required to ensure that $\langle \bar q q \rangle^0_\zeta$ is gauge
independent.  The parametrisation of Eq.~(\ref{SSM}) provides a model that
corresponds to the replacement $\gamma_m \to 1$ in Landau gauge, in which
case, with $S_0:=S_{u,m=0}$, Eq.~(\ref{condensate}) yields
\begin{eqnarray}
-\langle \bar u u \rangle_\zeta & = &
\lambda^3\,\frac{3}{4\pi^2}\,
\frac{b_0^u}{b_1^u\,b_3^u}\,\ln\frac{\zeta^2}{\Lambda_{\rm QCD}^2}\,.
\end{eqnarray}
This is a signature of DCSB in the model.

In Ref.~\cite{mishaB} the parameters $\bar m_f$, $b_{0\ldots 3}^f$ in
Eqs.~(\protect\ref{SSM}) and (\protect\ref{SVM}) take the values
\begin{equation}
\label{tableA} 
\begin{array}{c|lllll}
        & \;\bar m_f& b_0^f & b_1^f & b_2^f & b_3^f \\\hline
 u\;  & \;0.00897 & 0.131 & 2.90 & 0.603 & 0.185 \\
 s\;  & \;0.224   & 0.105 & 2.90 & 0.740 & 0.185
\end{array}
\end{equation}
with $\lambda=0.566\,$GeV, which were determined~\cite{brtpik} in a
least-squares fit to a range of light-hadron observables, and we note that
with $\Lambda_{\rm QCD}=0.2\,$GeV they yield $\langle \bar u u
\rangle_{1\,{\rm GeV}^2}= (-0.22\,{\rm GeV})^3$ and $\langle \bar s s
\rangle_{1\,{\rm GeV}^2}= 0.8\,\langle \bar u u \rangle_{1\,{\rm GeV}^2}$.
Herein we reconsider this parametrisation and allow $m_{u,s}$, $b_1^{u,s}$
and $b_2^{u,s}$ to vary. This is a reasonable step provided that in re-fitting
to an increased sample of observables the light-quark propagators are
pointwise little changed.

\subsection{Bethe-Salpeter amplitudes}
\subsubsection{Light Pseudoscalar Mesons}
The light-meson Bethe-Salpeter amplitudes can be determined reliably by
solving the Bethe-Salpeter equation (BSE) in a truncation consistent with
that employed in the quark DSE~\cite{mr97}.  However, since for the
light-quarks we have parametrised the solution of the quark DSE, we follow
Refs.~\cite{pipi,cdrpion,brtpik,mrpion,photopi,pichowsky,ph98,mishaB,mishaPLB}
and do the same for the light-meson amplitude; i.e., for the $\pi$- and
$K$-mesons we employ $\Gamma_{\pi,K}(k;P)= i\gamma_5\,{\cal E}_{\pi,K}(k^2)$
with,
\begin{equation} 
\label{piKamp}
{\cal E}_{P}(k^2)  =  \frac{1}{\hat f_{P}}\,B_{P}(k^2)\,,\;P=\pi,K\,,
\end{equation}
$\hat f_\pi= f_\pi/\surd 2$, where $B_P:=\left.B_u\right|_{b_0^u\to b_0^P}$,
obtained from Eq.~(\ref{defS}), and $b_0^{\pi,K}$ are allowed to
vary.\footnote{
In the following we explicitly account for the flavour structure in the
hadronic tensors.  With Eq.~(\protect\ref{piKamp}) we correct an error in
Eq.~(33) of Ref.~\protect\cite{mishaB}, which led to $\lsim 10\,$\%
underestimates of $f_+^{B\pi}(0)$, $f_+^{D K}(0)$ and $f_+^{D\pi}(0)$.  A
corrected Table~I, accounting also for a factor of $\sqrt{2}$ arising through
a mismatch between the normalisation conventions for light- and heavy-mesons,
is obtained with $E=0.698\,$GeV and $\Lambda = 1.273\,$GeV, and yields
$\Sigma^2/N=0.59$ cf. $0.48$ therein.}
This {\it Ansatz} follows from the constraints imposed by the axial-vector
Ward-Takahashi identity and, together with Eqs.~(\ref{SSM}) and (\ref{SVM}),
provides an algebraic representation of $\chi_P(k;p)$ valid for small to
intermediate meson energy~\cite{mrpion,mishaB}.

With this representation, Eqs.~(\ref{massP}) and (\ref{rP}) yield the
following expression for the $\pi$- and $K$-meson masses:
\begin{equation}
\hat f_{P}^2 \, m_P^2 = -( m_u +  m_{f^P})\,
\langle \bar q q\rangle_P^{1\,{\rm GeV}^2}\,,
\end{equation}
where $m_{f^\pi}=m_d$, $m_{f^K}=m_s$, and the ``in-meson condensate'' is
\begin{equation}
\langle \bar q q\rangle_P^{1\,{\rm GeV}^2}=
        -\,\lambda^3\,\ln\frac{1}{\Lambda_{\rm QCD}^2}\,\frac{3}{4\pi^2}\,
\frac{b_0^P}{b_1^u\,b_3^u}\,,\;P=\pi,K\,,
\end{equation}
and takes typical values~\cite{mr97}
$\langle \bar q q\rangle_\pi^{1\,{\rm GeV}^2}= 
1.05\,\langle \bar u u \rangle_{1\,{\rm GeV}^2}$
and
$\langle \bar q q\rangle_K^{1\,{\rm GeV}^2}= 
1.64\,\langle \bar u u \rangle_{1\,{\rm GeV}^2}$.

\subsubsection{Light Vector Mesons}
\label{seclvv}
The application of DSE-based phenomenology to processes involving vector
mesons is less extensive than that involving pseudoscalars.  Therefore the
modelling of vector meson Bethe-Salpeter amplitudes is less sophisticated.
Solutions of a mutually consistent truncation of the quark DSE and meson BSE;
e.g., Ref.~\cite{jm93}, indicate that a given vector meson is narrower in
momentum space than its pseudoscalar partner but that for both vector and
pseudoscalar mesons this width increases with the total current-mass of the
constituents.  These observations are confirmed in calculations of vector
meson electroproduction cross sections~\cite{pichowsky} and electromagnetic
form factors~\cite{ph98}.  The simple {\it Ansatz}
\begin{equation}
\label{gammaV}
\Gamma^V_{\mu}(k;p) = \frac{1}{{\cal N}^V}\,
\left(\gamma_\mu + p_\mu\,\frac{\gamma\cdot p}{M_V^2}\right)\,
\varphi(k^2)\,,
\end{equation}
where $\varphi(k^2) = 1/(1+k^4/\omega_V^4)$ with $\omega_V$ a parameter and
${\cal N}^V$ fixed by Eq.~(\ref{vnorm}), allows for the realisation of these
qualitative features and, from Ref.~\cite{ph98}, we expect $\omega_{K^\ast}
\approx 1.6\,\omega_\rho$.

\subsubsection{Heavy Mesons}
Renormalisation-group-improved ladder-like truncations of $K$ employed, e.g.,
in Refs.~\cite{mr97}, are inadequate for heavy-mesons; one reason being that
they do not yield the Dirac equation when the mass of one of the fermions
becomes large.  While an improved truncation valid in this regime is being
sought, there is currently no satisfactory alternative and the ladder-like
truncations have been used in spite of their inadequacy~\cite{jm93,conrad}.
Such studies cannot yield a complete and quantitatively reliable spectrum,
however, the result that heavy mesons are described by an amplitude whose
width behaves as described in Sec.~\ref{seclvv} must be qualitatively robust.
We therefore use a simple model for the amplitudes that allows a
representation of this feature: Eq.~(\ref{gammaV}) for heavy vector mesons,
with $\varphi(k^2)\to \varphi_H(k^2)$, and its analogue for heavy
pseudoscalar mesons
\begin{equation}
\label{HPamp}
\Gamma_P(k;p)= \frac{1}{{\cal N}^P}\,i\,\gamma_5\,
\varphi_H(k^2)\,,
\end{equation}
where $\varphi_H(k^2) = \exp\left(-k^2/\omega_H^2\right)$ and the
normalisation is fixed by Eq.~(\ref{psnorm}).  We assume the widths are spin
and flavour independent; i.e., $\omega_B = \omega_{B^\ast}= \omega_{B_s}$,
etc., as would be the case in the limit of exact heavy-quark symmetry, which
is a useful but not necessary simplification.

\section{Other Decay Processes}
\label{sece}
\subsection{Radiative Decays}
The radiative decays: $H^\ast(p_1) \to H(p_2)\,\gamma(k)\,,$ where $H=
D_{(s)}$, $B_{(s)}$, are described by the invariant amplitude
\begin{eqnarray}
\label{radiate}
\lefteqn{A(H^\ast \to H\,\gamma) = 
\epsilon_\mu^{\lambda_H}(p_1)\,
\epsilon_\nu^{\lambda_\gamma}(k) 
\left[ e_Q\,M_{\mu\nu}^Q(p_1,p_2) + e_q\,M_{\mu\nu}^q(p_1,p_2)\,
\right]}\\
& = & \varepsilon_{\mu\nu\alpha\beta}\,
\epsilon_\mu^{\lambda_H}(p_1)\,
\epsilon_\nu^{\lambda_\gamma}(k) \,
p_{1\alpha}\,k_{\beta}\left[e_Q\,J^Q(t) + e_q\,J^q(t)\right]\,,
\end{eqnarray}
where $t=-k^2=-(p_1-p_2)^2=0$ and $e_f$ is the fractional charge of the
active quark in units of the positron charge.  The sum indicates that the
decay occurs via a spin-flip transition by either the heavy or light quark.
The width is
\begin{equation}
\label{radiative}
\Gamma_{H^\ast \to H\gamma}=
\frac{\alpha_{\rm em}}{24 M_{H^\ast}^3}\,\lambda^{3/2}(M_{H^\ast}^2,m_H^2,0)\,
        \left[e_Q\,J^Q(0) + e_q\,J^q(0)\right]^2\,.
\end{equation}

In impulse approximation the hadronic tensors in Eq.~(\ref{radiate}) are:
\begin{eqnarray}
\nonumber \lefteqn{ {\cal M}^{Q}_{\mu\nu}(p_1,p_2)  = }\\
&& 2N_c\,{\rm tr}_D\,\int_\ell^\Lambda\,
\bar\Gamma_P(\ell;-p_2)\,S_{Q}(\ell_2) \,
i\Gamma_\nu^Q(\ell_2,\ell_1)\,
S_Q(\ell_1)\,\Gamma^V_\mu(\ell;p_1)\,S_{q}(\ell),
\label{iaradQ}\\
\nonumber \lefteqn{{\cal M}^{q}_{\mu\nu}(p_1,p_2)  = }\\
&&  2N_c\,{\rm tr}_D\,\int_\ell^\Lambda\,
\bar\Gamma_P(\ell;-p_2)\,S_{Q}(\ell_1) \,
\Gamma^V_\mu(\ell;p_1)\,S_{q}(\ell)\,i\Gamma_\nu^q(\ell,\ell+k)\,S_q(\ell+k)\,,
\end{eqnarray}
where $Q=c$ or $b$, $q=u$ or $s$ and $\ell_{1,2} = \ell+p_{1,2}$.

$\Gamma_\nu^f(\ell_1,\ell_2)$ is the dressed-quark-photon vertex, which
satisfies the vector Ward-Takahashi identity:
\begin{equation}
\label{vwti}
(\ell_1 - \ell_2)_\nu \, i\Gamma^f_\nu(\ell_1,\ell_2) = 
S_f^{-1}(\ell_1) - S_f^{-1}(\ell_2)\,,
\end{equation}
a feature that ensures current conservation~\cite{cdrpion}.
This vertex has been much studied~\cite{ayse97} and,
although its exact form remains unknown, its qualitatively robust features
have been elucidated so that a phenomenologically efficacious Ansatz has
emerged~\cite{bc80}:
\begin{eqnarray}
\nonumber \lefteqn{i\Gamma^f_\nu(\ell_1,\ell_2) = }\\ & & \label{bcvtx}
i\Sigma_A(\ell_1^2,\ell_2^2)\,\gamma_\mu +
(\ell_1+\ell_2)_\mu\,\left[\case{1}{2}i\gamma\cdot (\ell_1+\ell_2) \,
\Delta_A(\ell_1^2,\ell_2^2) + \Delta_B(\ell_1^2,\ell_2^2)\right]\,;\\
&& \Sigma_F(\ell_1^2,\ell_2^2) = \case{1}{2}\,[F(\ell_1^2)+F(\ell_2^2)]\,,\;
\Delta_F(\ell_1^2,\ell_2^2) =
\frac{F(\ell_1^2)-F(\ell_2^2)}{\ell_1^2-\ell_2^2}\,,
\end{eqnarray}
where $F= A_f, B_f$; i.e., the scalar functions in Eq.~(\ref{defS}).  A
feature of Eq.~(\ref{bcvtx}) is that the vertex is completely determined by
the dressed-quark propagator.  In Landau gauge the quantitative effect of
modifications, such as that canvassed in Ref.~\cite{cp92}, is small and can
be compensated for by small changes in the parameters that characterise a
given model study~\cite{hawes}.  The structure in Eq.~(\ref{bcvtx}) is only
important for light-quarks because, using Eq.~(\ref{dsehq}): $A_Q\equiv 1$,
$B_Q\equiv \hat M_Q$, and hence
\begin{equation}
\Gamma^Q_\nu(\ell_1,\ell_2) = \gamma_\mu\,.
\end{equation}

\subsection{Strong Decays}
The process $H^\ast(p_1) \to H(p_2)\, \pi(q)$,
with $p_1^2=-M_{H^\ast}^2$, $p_2^2=-m_H^2$ and $q^2=-m_\pi^2$, is described
by the invariant amplitude
\begin{equation}
A(H^\ast\to H\pi) = \epsilon_\mu^{\lambda_{H^\ast}}(p_1) \,
                M_\mu^{H^\ast H \pi}(p_1,p_2) 
:= \epsilon_\mu^{\lambda_{H^\ast}}(p_1)\,p_{2\mu}\,g_{H^\ast H \pi}\,.
\end{equation}
$g_{H^\ast H \pi}$ can be calculated even when the decay is kinematically
forbidden, as for $B^\ast$, and is sometimes re-expressed via
\begin{equation}
\bar g_{H^\ast H \pi} := \frac{f_\pi}{2 m_H} \,g_{H^\ast H \pi}\,.
\end{equation}
We calculate the coupling using the impulse approximation, Eq.~(\ref{HsHpi}),
and this gives the width:
\begin{equation}
\Gamma_{H^\ast H \pi} = \frac{g_{H^\ast H \pi}^2}{192\,\pi\,M_{H^\ast}^5}\,
                        \lambda^{3/2}(m_H^2,m_\pi^2,M_{H^\ast}^2)\,.
\end{equation}

We also consider the analogous decays of light vector mesons:
$\rho(P=k_1+k_2) \to \pi(k_1)\pi(k_2)$ and $K^\ast(P) \to K(k_1)\pi(k_2)$.
For these processes the hadronic current can be written
\begin{eqnarray}
M_\mu^{VP\pi}(k_1,k_2) & = & 
(k_1-k_2)_\mu\,f^+(t) + P_\mu\,f^-(t)\,,
\end{eqnarray}
which we calculate using the impulse approximation, Eq.~(\ref{VPpi}).  The
decay constant is
\begin{equation}
g_{VP\pi}= f^+(t=M_V^2)\,,
\end{equation}
in terms of which the widths are
\begin{eqnarray}
\Gamma_{\rho\pi\pi} &= &\frac{g_{\rho\pi\pi}^2}{48\pi\,M_\rho^5}\,
        \lambda^{3/2}(m_\pi^2,m_\pi^2,M_\rho^2)\,,\\ 
\Gamma_{K^\ast (K\pi)} &= &\frac{g_{K^\ast K\pi}^2}{64\pi\,M_{K^\ast}^5}\,
        \lambda^{3/2}(m_K^2,m_\pi^2,M_{K^\ast}^2)\,.
\end{eqnarray}

\subsection{Rare Flavour-Changing Neutral-Current Process}
The final decay we consider is the rare, flavour-changing neutral current
process: $B(p_1)\to K^\ast(p_2) \gamma(q)$, which proceeds
predominantly~\cite{buchalla} via the local magnetic penguin operator:
\begin{equation}
Q_{7\gamma}:= \frac{e}{8\pi^2}\,\hat M_b\,F_{\mu\nu}\,
        \bar s\,\sigma_{\mu\nu} (1 + \gamma_5) \,b\,,
\end{equation}
where $F_{\mu\nu}$ is the photon's field strength tensor, and the effective
interaction promoting this process, renormalised at a scale $\zeta \sim \hat
M_b$, is
\begin{equation}
H^{\rm eff}= -\frac{G_F}{\surd 2}\, V^\ast_{ts}V_{tb}\,C_7 \,Q_{7\gamma}\,,
\end{equation}
with $C_7(\hat M_b) = -0.299$ and $|V^\ast_{ts}V_{tb}|^2 = (0.95\pm
0.03)\,|V_{cb}|^2$.  Consequently, the invariant amplitude describing the
decay is
\begin{eqnarray}
A(B\to K^\ast \gamma) 
        & = & -\frac{G_F}{\surd 2}\, V^\ast_{ts}V_{tb}\,C_7 \,
        \frac{e \hat M_b}{4\pi^2}\,
        \epsilon_\mu^{\lambda_\gamma}(q)\,
        \epsilon_\nu^{\lambda_{K^\ast}}(p_2)\,
        M_{\mu\nu}^{B\to K^\ast \gamma}(p_1,p_2)\,, 
\end{eqnarray}
with, at $q^2=0$,
\begin{eqnarray}
\nonumber \lefteqn{\epsilon_\mu^{\lambda_\gamma}(q)\,
        M_{\mu\nu}^{B\to K^\ast \gamma}(p_1,p_2) =}\\
&&      \epsilon_\mu^{\lambda_\gamma}(q)\,
        g_{B K^\ast \gamma}\,
        \left( \varepsilon_{\mu\nu\alpha\beta}\,
        p_{1\alpha}\,p_{2\beta} + 
        \delta_{\mu\nu} (m_B^2-M_{K^\ast}^2) - 
        (p_1+p_2)_\mu \,p_{1\nu} \right) .
\end{eqnarray}
We calculate the coupling using the impulse approximation to the hadronic
tensor, Eq.~(\ref{rareB}), in terms of which the decay width is
\begin{equation}
\Gamma_{B\to K^\ast \gamma} = \alpha_{\rm em}(\hat M_b)\,m_B^3\,\left(1 -
\frac{M_{K^\ast}^2}{m_B^2}\right)^3\, \frac{G_F^2}{32\pi^4}\, \hat M_b^2
C_7^2\, |V_{ts}^\ast V_{tb}|^2\, g_{B K^\ast \gamma}^2\,.
\end{equation}

\section{Heavy-Quark Symmetry Limits}
\label{secf}
With algebraic representations of the dressed-quark propagators and
Bethe-Salpeter amplitudes the calculation of all observables is
straightforward.  In addition, one can obtain simple formulae that express
the heavy-quark symmetry limits.  We present them here, and in
Sec.~\ref{secg} gauge their accuracy and relevance through a comparison with
the results of our complete calculations.

\subsection{Leptonic Decays}
To begin, using Eq.~(\ref{hqf}), Eqs.~(\ref{psnorm}) and (\ref{vnorm}) with
$K$ assumed $p$-independent in order to effect a consistent impulse
approximation, and Eq.~(\ref{HPamp}) with its analogue for the heavy vector
mesons, one finds~\cite{mishaPLB}
\begin{eqnarray}
\label{kappaf}
\lefteqn{ \frac{1}{m_H}\,\frac{1}{\kappa_f^2} :=
 {\cal N}_P^2 = {\cal N}_V^2  =}\\
&& \frac{1}{m_H} \,\frac{N_c}{4\pi^2}\,
\int_0^\infty\,du\,\varphi^2_H(z)\, 
\left\{ \sigma_S^f(z) + \surd{u}\,\sigma_V^f(z)\right\}\,,
\end{eqnarray}
where $z=u-2 E_H \surd u$, $f$ labels the meson's lighter quark and in this
section all dimensioned quantities are expressed in units of our mass-scale,
$\lambda$.\footnote{ Here and in the following we employ methods analogous to
that described in the appendix of Ref.~\cite{mishaB} to simplify the
integrands that express observables.}
This illustrates Eq.~(\ref{calGP}).  Similarly, using this result with
Eq.~(\ref{hqf}) in Eqs.~(\ref{fwk}) and (\ref{fwkV}),
\begin{eqnarray}
\nonumber \lefteqn{f_P = f_V = }\\
&& \frac{1}{\surd m_H}\,\kappa_f\,\frac{N_c}{2\sqrt{2} \pi^2}\,
        \int_0^\infty\,du\,\left(\surd u - E_H\right)\,
        \varphi_H(z)\,
     \left\{ \sigma_S^f(z) + \case{1}{2}\surd{u}\,\sigma_V^f(z)\right\}\,,
\end{eqnarray}
which is the promised illustration of Eqs.~(\ref{fwkasymp}), (\ref{fwkVasymp})
and (\ref{fpvasymp}).

\subsection{Semileptonic Heavy$\,\to\,$Heavy Transitions}
The semileptonic heavy$\,\to\,$ heavy transition form factors acquire a
particularly simple form in the heavy-quark symmetry limit.  From
Eqs.~(\ref{fpfm}) and (\ref{ia}) one obtains~\cite{mishaPLB}:
\begin{eqnarray}
\label{fxi}
\lefteqn{f_\pm(t)= {\cal T}_\pm \,\xi_f(w)
:= \frac{m_{P_2} \pm 
m_{P_1}}{2\sqrt{m_{P_2}m_{P_1}}}\,\xi_f(w),}\\
&& \label{xif}
\xi_f(w)  =  \kappa_f^2\,\frac{N_c}{4\pi^2}\,
\int_0^1 d\tau\,\frac{1}{W}\,
\int_0^\infty du \, \varphi_H(z_W)^2\,
        \left[\sigma_S^f(z_W) + \sqrt{\frac{u}{W}} \sigma_V^f(z_W)\right]\,,
\end{eqnarray}
with $\kappa_f$ defined in Eq.~(\ref{kappaf}), $W= 1 + 2 \tau (1-\tau)
(w-1)$, $z_W= u - 2 E_H \sqrt{u/W}$ and
\begin{equation}
w = \frac{m_{P_1}^2 + m_{P_2}^2 - t}{2 m_{P_1} m_{P_2}} = - v_{P_1} \cdot
v_{P_2}\,. 
\end{equation}
The canonical normalisation of the Bethe-Salpeter amplitude automatically
ensures that
\begin{equation}
\label{xione}
\xi_f(w=1) = 1
\end{equation}
and it follows~\cite{mishaPLB} from Eq.~(\ref{xif}) that
\begin{equation}
\rho^2 := -\left.\frac{d\xi_f}{dw}\right|_{w=1} \geq \frac{1}{3}\,.
\end{equation}

Similar analysis applied to Eqs.~(\ref{vsl}) and (\ref{ia}) yields
\begin{eqnarray}
\label{a1xi}
&& A_1(t)= \frac{1}{{\cal T}_+}\,\case{1}{2}\,(1+w)\,\xi_f(w)\\
\label{a2xi}
&& A_2(t) = -A_3(t) = V(t) = {\cal T}_+\,\xi_f(w)\,.
\end{eqnarray}
Equations~(\ref{fxi}), (\ref{a1xi}) and (\ref{a2xi}) are exemplars of a
general result that, in the heavy-quark symmetry limit, the semileptonic $H_f
\to H_f^\prime$ transitions are described by a single, universal function:
$\xi_f(w)$~\cite{IW90}.  In this limit the functions
\begin{eqnarray} 
\label{Rratios}
R_1(w) &:= & (1 - t/t_+) \,\frac{V(t)}{A_1(t)}\,,\\
\label{Rratiosb}
R_2(w) &:= & (1 - t/t_+) \,\frac{A_2(t)}{A_1(t)}
\end{eqnarray}
are also constant ($=1$), independent of $w$.

\subsection{Semileptonic Heavy$\,\to\,$Light Transitions}
In this case the form factors cannot be expressed in terms of a single
function when the heavy meson mass becomes large.  However, some
simplifications do occur.  Adapting the analysis employed above,
Eqs.~(\ref{fpfm}) and (\ref{ia}) yield
\begin{eqnarray}
\nonumber \lefteqn{f_\pm(t)  = }\\
&&  \label{fpmhq} \frac{N_c \kappa_f}{4\pi^2}
\int_0^\infty\,du\,\left(\surd u -E_H\right)\,
\varphi_H(z)\,{\cal E}_P(z)\,
\int_0^1\,d\tau\,\surd \hat M_Q\,
\left[\frac{1}{\hat M_Q} H_1 \pm H_2\right],
\end{eqnarray}
where $z_1= z+2 X \tau \surd u$, $X= (\hat M_Q/2)\,(1 - t/\hat M_Q^2)$, $z_2=
(1-\tau) (z + \tau u)$, with similar expressions for $A_1,A_2,A_3$ and $V$,
Eqs.~(\ref{a1hq})-(\ref{vhq}), and the functions forming the integrands, $H =
H(z,z_1,z_2)$, also given in the appendix.

When $E_H/m_H \to 0$ and $\hat M_Q\to m_H\to\infty$ then $X=0$ at the end
point, $t=t_{\rm max}$, and one obtains the following simple scaling
laws~\cite{IW90}:
\begin{eqnarray}
\label{scala}
\frac{f^{P_1}_+ + f^{P_1}_-}{f^{P_2}_+ + f^{P_2}_-}
= \frac{A^{P_1}_2 + A^{P_1}_3}{A^{P_2}_2 + A^{P_2}_3} = 
\frac{A^{P_1}_1}{A^{P_2}_1} & =  &
 \sqrt{\frac{m_{P_2}}{m_{P_1}}} \,,
\\
\label{scalb}
\frac{f^{P_1}_+ - f^{P_1}_-}{f^{P_2}_+ - f^{P_2}_-} = 
\frac{A^{P_1}_2 - A^{P_1}_3}{A^{P_2}_2 - A^{P_2}_3} = 
\frac{V^{P_1}}{V^{P_2}} & = &  \sqrt{\frac{m_{P_1}}{m_{P_2}}}\,.
\end{eqnarray}
It also clear from Eqs.~(\ref{fpmhq}) and (\ref{a1hq})-(\ref{vhq}) that in
this limit
\begin{equation}
f_-(t_{\rm max})=- f_+(t_{\rm max})\,,\;
A_3(t_{\rm max})= - A_2(t_{\rm max})
\end{equation}
and hence one obtains a further idealisation from Eqs.~(\ref{scalb}):
\begin{eqnarray}
\label{hlIdeal}
\frac{f^{P_1}_+ }{f^{P_2}_+} = 
\frac{A^{P_1}_2}{A^{P_2}_2} & = & \sqrt{\frac{m_{P_1}}{m_{P_2}}}\,.
\end{eqnarray}

The form factors also exhibit a factorisable mass-dependence at $t=0$ but it
is modulated by the properties of the light-quark propagator; e.g.,
\begin{eqnarray}
\nonumber \lefteqn{f_+(0)  = }\\
&&  \label{fpt0hq} \frac{\ln  m_H}{ m_H^{1/2}}\,
\frac{N_c \kappa_f}{4\pi^2}
\int_0^\infty\,du\,\left(1 - E_H/\surd u \right)\,
\varphi_H(z)\,{\cal E}_P(z)\,
\left\{ m_f \sigma^f_S(z) + z \sigma^f_V(z) \right\},
\end{eqnarray}
with the forms of $A_1,A_2,A_3$ and $V$ given in Eq.~(\ref{otherst0hq}). The
factorised dependence on $\ln m_H/ m_H^{1/2}$ is a common feature of our
impulse approximation to all heavy$\,\to\,$light form factors and differs
from the $1/m_H^{3/2}$-behaviour obtained in LCSRs~\cite{braun}.  It arises
in the simplification of the multidimensional integrals because our model
dressed-light-quark propagators evolve into free-particle propagators at
large spacelike momenta.  In QCD, where the propagators and Bethe-Salpeter
amplitudes possess an anomalous dimension, we expect only that
$\ln m_H \to \ln^\gamma m_H$, where $\gamma$ is calculable.

\subsection{Other Decay Processes}
\label{raddecays}
In the heavy-quark symmetry limit Eq.~(\ref{radiative}) can be expressed in a
particularly simple form because
\begin{equation}
J^Q(0) = \frac{1}{\hat M_Q}\,,
\end{equation}
which suggests the following ``classical'' formulation of the width
\begin{equation}
\label{radiative2}
\Gamma_{H^\ast \to H\gamma}= \frac{\alpha_{\rm em}}{6
M_{H^\ast}^3}\,\lambda^{3/2}(M_{H^\ast}^2,m_H^2,0)\, \left(\mu_Q +
\mu_q\right)^2\,,
\end{equation}
where we have introduced the quark magnetons:
\begin{eqnarray}
&& \label{Mqeff} \mu_f =  \frac{e_f}{2\, M_f^{\rm in}} \,,\;
M_f^{\rm in}:= \frac{1}{J^f(0)}\,.
\end{eqnarray}
In the heavy-quark symmetry limit, $M^{\rm in}_Q = \hat M_Q$ and provides a
measure of the quark's {\it inertial} mass.

The expression for the $H^\ast \to H \pi$ coupling also simplifies:
\begin{eqnarray}
\nonumber \lefteqn{g_{H^\ast H \pi} 
= m_H\, \frac{\kappa_f N_c}{\sqrt{2}\pi^2}
\int_0^\infty\,du\,\left(\surd u - E_H\right)\,
        \varphi_H(z)^2\,{\cal E}_P(z)\,\times
}\\
&& \;\;\;\;\;\;\;\;\;\;\;\;\;\;\;
        \left\{\sigma_V \left[\sigma_S + \case{1}{2} \sigma_V \right]
        + \left(\case{2}{3}u - E_H \surd u\right)
        \left[\sigma_S\sigma_V^\prime 
                - \sigma_V\sigma_S^\prime\right]\right\}\,,
\end{eqnarray}
exposing a linear increase with the mass of the heavy meson.

\section{Results}
\label{secg}
The calculation of all observables is a straightforward numerical exercise.
We simplify the integrands using the methods illustrated in the appendix of
Ref.~\cite{mishaB} and the expressions we actually evaluate are similar to
those presented therein with the important difference, however, that herein
we use the constituent-like dressed-quark propagator of Eq.~(\ref{dsehq}) for
the $c,b$-quarks, {\it not} Eq.~(\ref{hqf}).

Our model has ten parameters plus the four quark masses, all of which are
fixed via a $\chi^2$-fit to the $N_{\rm obs}=42$ heavy- and light-meson
observables in Tables~\ref{tableC} and~\ref{tableD}.  This yields\footnote{
In the fitting we used~\cite{pdg98}: $V_{ub}=0.0033$, $V_{cd}= 0.2205$,
$V_{cs}= 0.9745$ and $V_{cb}= 0.039$; and, in GeV, $M_\rho= 0.77$,
$M_{K^\ast}=0.892$ and, except in the kinematic factor
$\lambda(m_1^2,m_2^2,t)$ where the splittings are crucial, averaged $D$- and
$B$-meson masses: $m_D= 1.99$, $m_B=5.35$ (from $m_D=1.87$, $m_{D_s}=1.97$,
$M_{D^\ast}=2.01$, $M_{D_s^\ast}=2.11$, and $m_B=5.28$, $m_{B_s}=5.37$,
$M_{B^\ast}=5.32$, $M_{B_s^\ast}=5.42$).}
\begin{equation}
\label{tableB} 
\begin{array}{c|lll}
      & \;\bar m_f& b_1^f & b_2^f\\\hline
 u\;  & \;0.00948 & 2.94 & 0.733 \\
 s\;  & \;0.210   & 3.18 & 0.858 
\end{array}\;\;
\begin{array}{c|l}
    & b_0^P \\ \hline
\pi & 0.204 \\
K   & 0.319
\end{array}\;\;
\begin{array}{c|l}
      & \omega_V^{\rm GeV} \\ \hline
\rho  & 0.515 \\
K^\ast& 0.817
\end{array}\;\;
\begin{array}{c|l}
      & \omega_H^{\rm GeV} \\ \hline
   D  & 1.81 \\
   B  & 1.81
\end{array}\;\;
\begin{array}{c|l}
      & \hat M_Q^{\rm GeV} \\ \hline
   c  & 1.32 \\
   b  & 4.65
\end{array}
\end{equation}
with $\chi^2/{\rm d.o.f} = 1.75$ and $\chi^2/N_{\rm obs} = 1.17$.  The
dimensionless $u,s$ current-quark masses correspond to $m_u = 5.4\,$MeV and
$m_s=119\,$MeV, and the parameters $b_0^{u,s}$ and $b_3^{u,s}$, which were
not varied, are given in Eq.~(\ref{tableA}).  A general comparison of
Eq.~(\ref{tableA}) with Eq.~(\ref{tableB}) reveals the light-quark
propagators to be little modified, with an average change in the parameters
of only 3\%, and with this parameter set
\begin{equation}
\label{usME}
M^E_u= 0.36 \,{\rm GeV}\,,\;M^E_s= 0.49\,{\rm GeV}\,.
\end{equation}
These values are little changed from those obtained with Eq.~(\ref{tableA}):
$M^E_u= 0.35\,,\;M^E_s= 0.48$, and are similar to the constituent-light-quark
masses typically employed in quark models~\cite{simon}: $M_{u/d}=0.33\,$GeV
and $M_s=0.55\,$GeV.  Further, $\omega_{K^\ast}/\omega_\rho=1.59$, which is
identical to the value calculated from Ref.~\cite{ph98}, as we anticipated in
Sec.~\ref{seclvv}, and our simple parametrisation yields~\cite{privatemike}:
\begin{equation}
r_\rho= 0.69\,{\rm fm}, \;
\mu_\rho=2.44\,\case{e_0}{2M_\rho},
\end{equation}
values of the $\rho$-meson charge radius and magnetic moment that compare
well with those in Ref.~\cite{ph98}: $0.61\,$fm and $2.69\,e_0/(2M_\rho)$.

For the heavy quarks we note that their fitted masses are consistent with the
estimates of Ref.~\cite{pdg98} and hence that the heavy-meson binding energy
is large:
\begin{equation}
E_D:= m_D - \hat M_c = 0.67\,{\rm GeV}\,,\;
E_B:= m_B - \hat M_b = 0.70\,{\rm GeV}\,.
\end{equation}
$E_B$ calculated here is identical to the corrected heavy-meson binding
energy obtained using the heavy-quark symmetry methods of Ref.~\cite{mishaB},
while $E_D$ is $4$\% less.  These values yield $E_D/\hat M_c= 0.51$ and
$E_B/\hat M_b= 0.15$, which furnishes another indication that while a
heavy-quark expansion will be accurate for the $b$-quark it will provide a
poor approximation for the $c$-quark.  This is emphasised by the value of
$\omega_D = \omega_B$, which means that the Compton wavelength of the
$c$-quark is greater than the length-scale characterising the bound state's
extent.

Having fixed the model's parameters, in Tables~\ref{tableE} -~\ref{tableG} we
report the calculated values of a wide range of quantities not included in
the fitting.  Many articles report a calculation of some of these quantities
and Refs.~\cite{richman,gHsHpi,mishaC,simula} can be consulted for tabulated
comparisons.

\begin{figure}[t]
\centering{\ \epsfig{figure=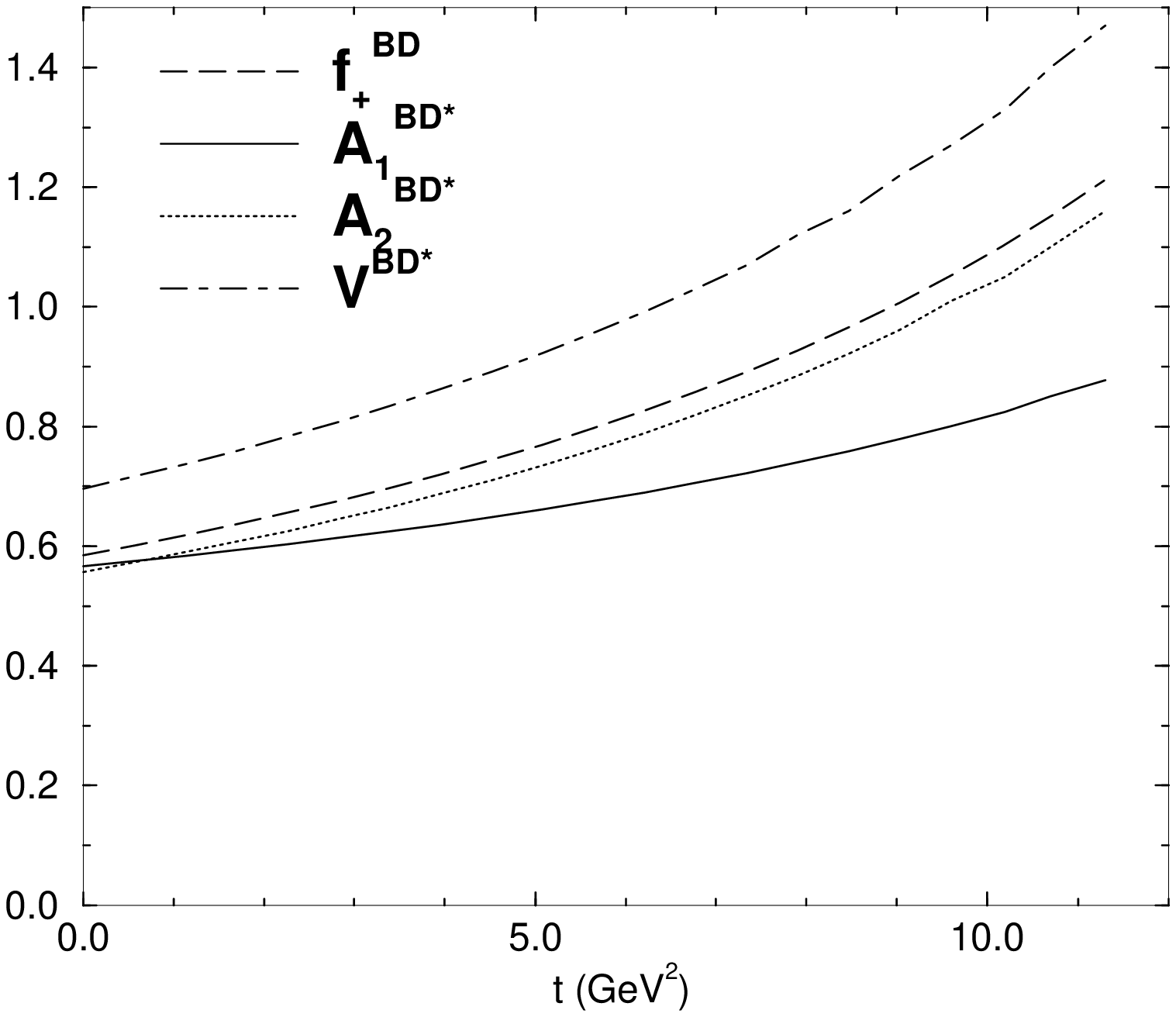,height=8.5cm}}\vspace*{1.0em}

\centering{\ \epsfig{figure=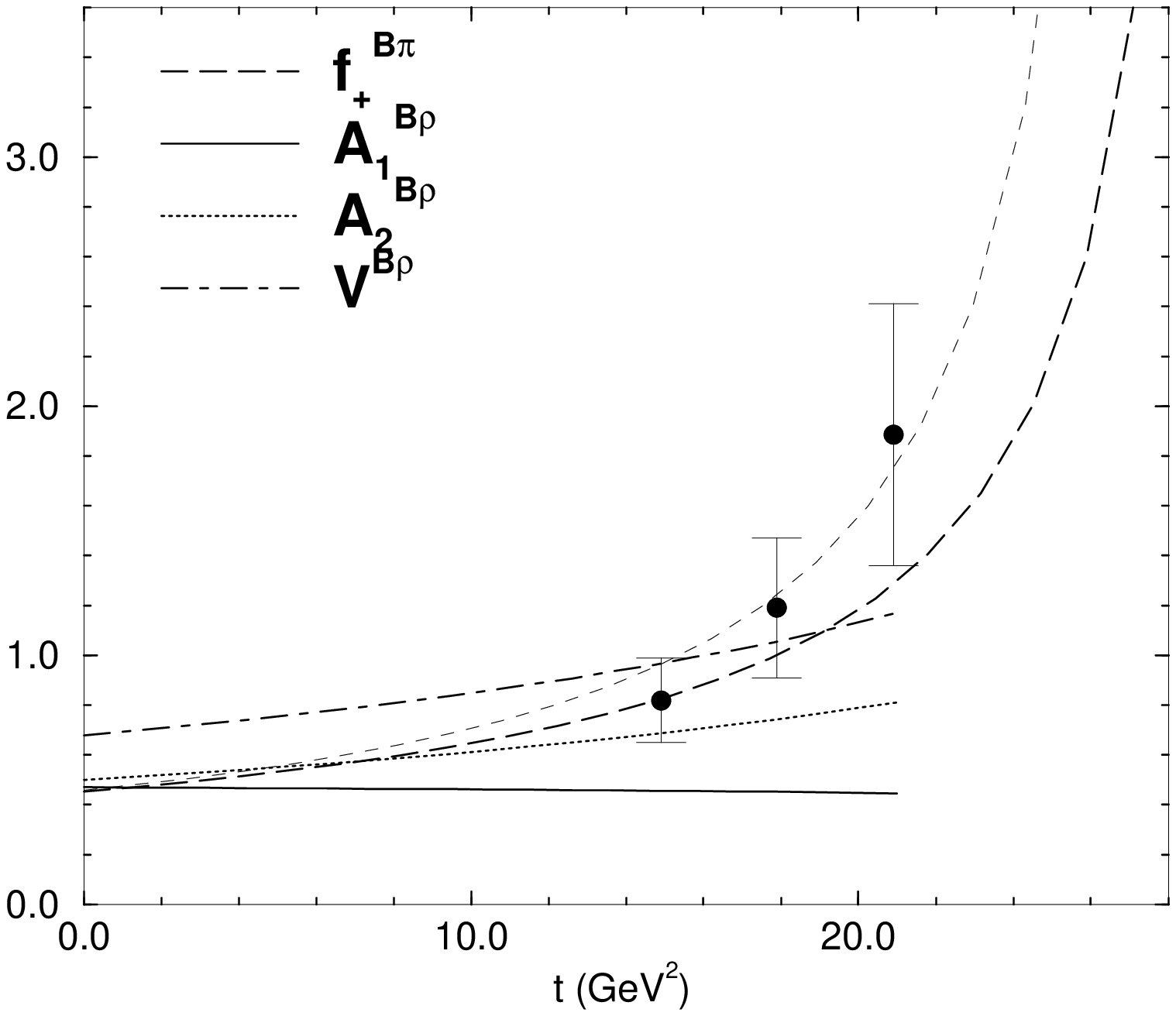,height=8.5cm}}
\caption{\label{figa} Upper panel: calculated form of the semileptonic $B\to
D$ and $B\to D^\ast$ form factors.  Lower panel: the semileptonic $B\to \pi$
and $B\to \rho$ form factors with, for comparison, data from a lattice
simulation~\protect\cite{latt} and a vector dominance, monopole model:
$f_+^{B\to \pi}(t)= 0.46/(1-t/m_{B^\ast}^2)$, $m_{B^\ast} = 5.325\,$GeV, the
light, short-dashed line.  Monopole fits to our calculated results are given
in Eqs.~(\protect\ref{mono}) and (\protect\ref{monomass}).}
\end{figure}

\begin{figure}[t]
\centering{\ \epsfig{figure=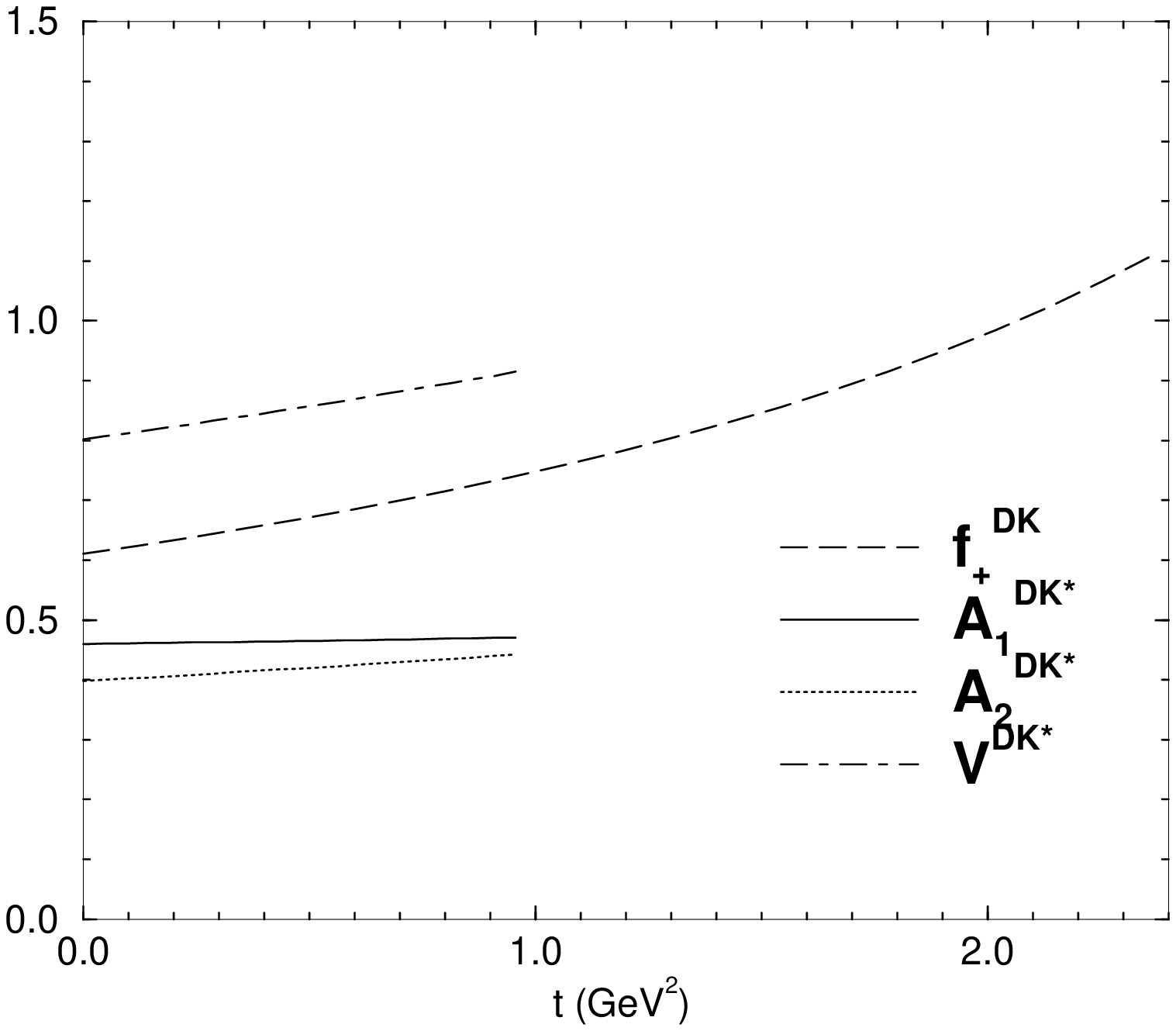,height=8.5cm}}\vspace*{1.0em}

\centering{\ \epsfig{figure=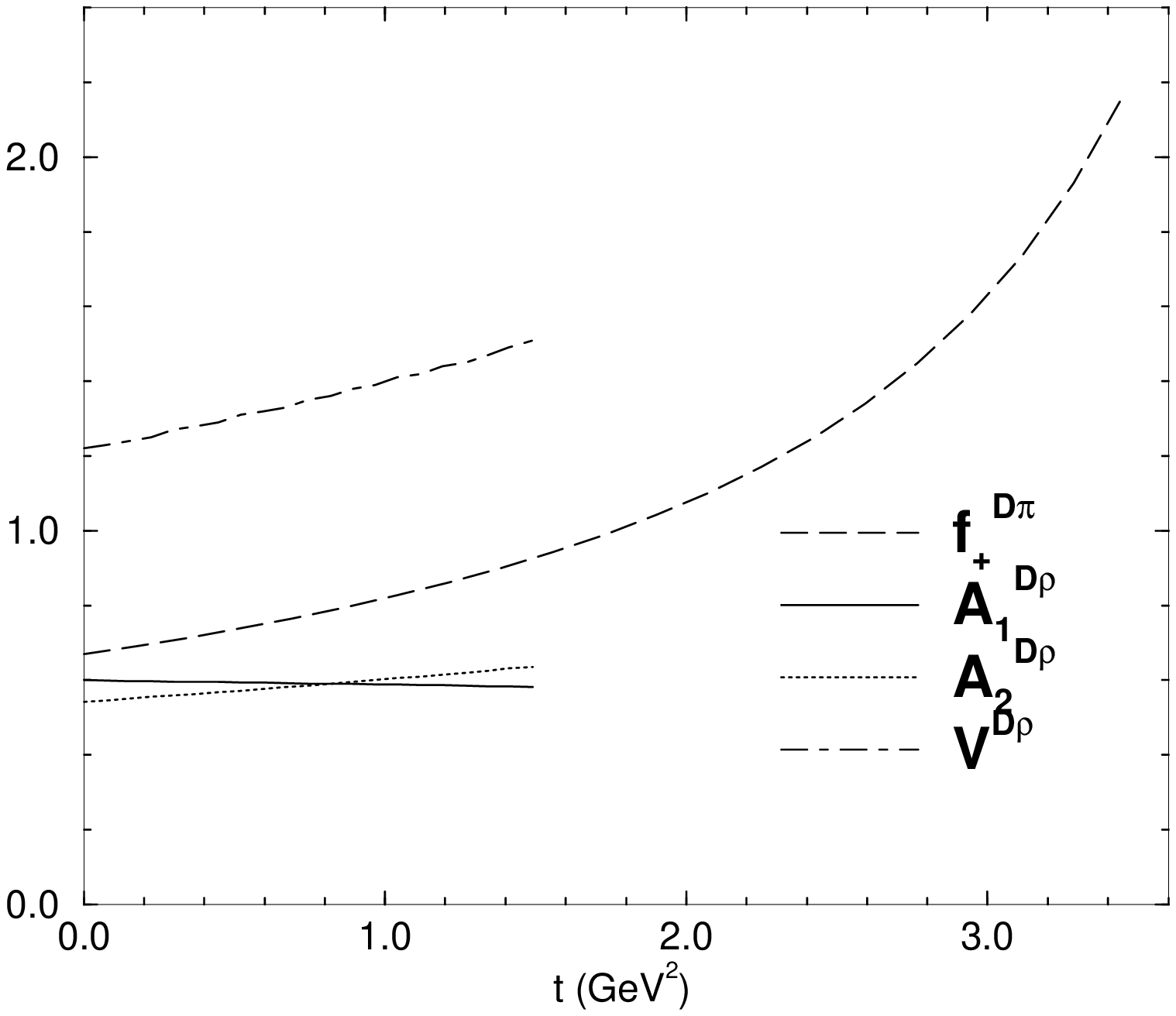,height=8.5cm}}
\caption{\label{figb} Calculated form of the semileptonic $D\to K$ and $D\to
K^\ast$ (upper panel), and $D\to \pi$ and $D\to \rho$ (lower panel) form
factors.  Monopole fits to our calculated results are given in
Eqs.~(\protect\ref{mono}) and (\protect\ref{monomass}).}
\end{figure}

The calculated $t$-dependence of the semileptonic transition form factors
that are the hadronic manifestation of the $b\to c$, $b\to u$, $c\to s$ and
$c\to d$ transitions is depicted in Figs.~\ref{figa} and~\ref{figb}.  The
form factors can be {\it approximated} by the monopole form
\begin{equation}
\label{mono}
h(t) = \frac{h(0)}{1 - t/h_1}\,,
\end{equation}
with the dimensionless values of $h(0)$ given in Tables~\ref{tableF}
and~\ref{tableG}, and $h_1$, in GeV$^2$, listed in Eq.~(\ref{monomass}).
\begin{equation}
\label{monomass}
\begin{array}{l|rrrr}
        & h_1^{f_+} & h_1^{A_1} & h_1^{A_2} & h_1^{V}\\\hline
B\to D, D^\ast& (4.63)^2 & (5.73)^2 & (4.64)^2 & (4.61)^2\\
B\to \pi,\rho & (5.58)^2 & -(21.5)^2 & (6.94)^2 & (7.06)^2\\
D\to K,K^\ast & (2.31)^2 & (6.70)^2 & (3.09)^2 & (2.78)^2 \\
D\to \pi,\rho & (2.25)^2 & -(7.06)^2 & (3.17)^2 & (2.80)^2
\end{array}
\end{equation}
As expected, in each of the channels the magnitude of the monopole mass is
determined primarily by the heavy-quark mass, with the actual value
reflecting small modifications due to the light-quark degrees of freedom and
bound state structure.  The behaviour of all the form factors is consistent
with lattice simulations, where results are available~\cite{flynn}.

\subsection{Fidelity of Heavy-Quark Symmetry}
\begin{figure}[t]
\centering{\ \epsfig{figure=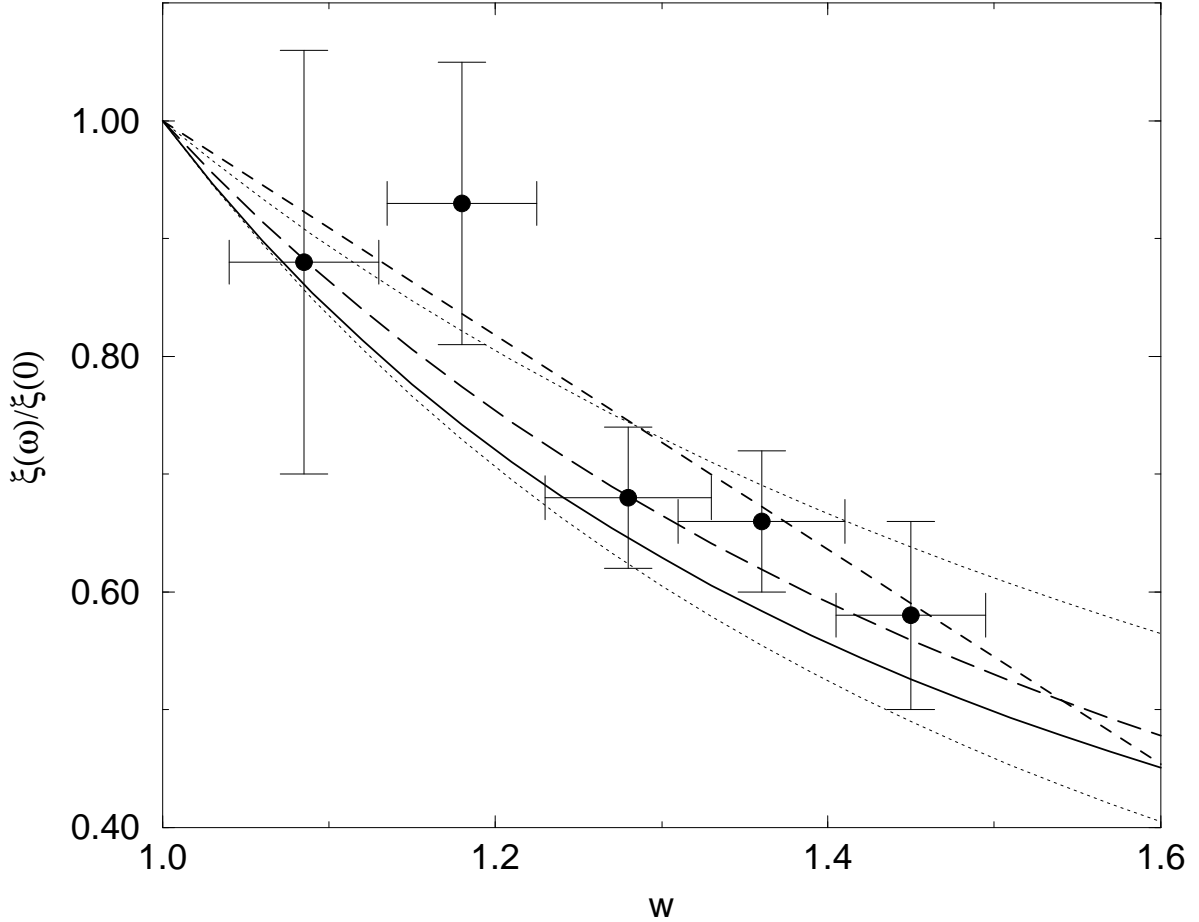,height=12.5cm}}
\caption{\label{fige} Calculated form of $\xi(w)$ compared with recent
experimental analyses.  Our result: solid line.  Experiment: short-dashed
line, linear fit from Ref.~\protect\cite{cesr96},
Eq.~(\protect\ref{cesrlinear}); long-dashed line, nonlinear fit from
Ref.~\protect\cite{cesr96}, Eq.~(\protect\ref{cesrnonlinear}).  The two
light, dotted lines are this nonlinear fit evaluated with the extreme values
of $\rho^2$: upper line, $\rho^2= 1.17$ and lower line, $\rho^2=1.89$; data
points, Ref.~\protect\cite{argus93}.}
\end{figure}

The universal function characterising semileptonic transitions in the
heavy-quark symmetry limit, $\xi(w)$, can be obtained most reliably from
$B\to D,D^\ast$ transitions, if it can be obtained at all.  Using
Eq.~(\ref{fxi}) to extract it from $f_+^{B\to D}(t)$ we obtain
\begin{equation}
\label{xifp}
\xi^{f_+}(1)= 1.08\,,
\end{equation}
which is a measurable deviation from Eq.~(\ref{xione}).  We plot
$\xi^{f_+}(w)/\xi^{f_+}(0)$ in Fig.~\ref{fige} and compare it with two
experimental fits~\cite{cesr96}:
\begin{eqnarray}
\xi(w) & = & 1 - \rho^2\,( w - 1), \; \rho^2 = 0.91\pm 0.15 \pm 0.16\,,
\label{cesrlinear}\\
\xi(w) & = & \frac{2}{w+1}\,\exp\left[(1-2\rho^2) \frac{w-1}{w+1}\right],
        \;\rho^2 = 1.53 \pm 0.36 \pm 0.14\,.
\label{cesrnonlinear}
\end{eqnarray}
The agreement obtained here is possible because, unlike Ref.~\cite{mishaB},
we did not employ a heavy-quark expansion for the $c$-quark.  Our calculated
result is well approximated by
\begin{equation}
\label{fitIW}
\xi^{f_+}(w) = \frac{1}{1 + \tilde\rho^2_{f_+}\,(w-1)  }\,,
\;\tilde\rho^2_{f_+}=1.98\,.
\end{equation}

We have also used Eqs.~(\ref{a1xi}) and (\ref{a2xi}) to extract $\xi(w)$ from
$B\to D^\ast$.  This yields
\begin{equation}
\xi^{A_1}(1)= 0.987\,,\;\xi^{A_2}(1)=1.03\,,\;
\xi^{V}(1)= 1.30\,,
\end{equation}
an $w$-dependence well-described by Eq.~(\ref{fitIW}) but with
\begin{equation}
\tilde\rho^2_{A_1}=1.79\,,\;
\tilde\rho^2_{A_2}=1.99\,,\;
\tilde\rho^2_{V}=2.02\,,
\end{equation}
and the ratios, Eqs.~(\ref{Rratios}) and (\ref{Rratiosb}), 
\begin{equation}
R_{1}(1)/R_{1}(w_{\rm max})=1.08\,,\;R_{2}(1)/R_{2}(w_{\rm max})=1.06\,.
\end{equation}
This collection of results furnishes a measure of the degree to which
heavy-quark symmetry is respected in $b\to c$ processes.  Combining them it
is clear that even in this case, which is the nearest contemporary
realisation of the heavy-quark symmetry limit, corrections of $\lsim 30$\%
must be expected.

The scaling laws in Eqs.~(\ref{scala}) and (\ref{scalb}), which relate the
heavy$\,\to\,$ light form factors at $t=t_-$, can be tested in
$B,D\to\pi,\rho$ decays and we find
\begin{equation}
\begin{array}{c|ccccccc}
\sqrt{\frac{m_D}{m_B}}=0.61 & \frac{A_2^B+A_3^B}{A_2^D+A_3^D} & 
       \frac{A_1^B}{A_1^D} & \frac{A_2^D-A_3^D}{A_2^B-A_3^B} &
        \frac{V^D}{V^B} &  \frac{A_2^D}{A_2^B} & 
        \frac{A_3^D}{A_3^B} & \frac{f_+^D}{f_+^B} \\ \hline 
\frac{E_D}{\hat M_c}= 0.51,\,
\frac{E_B}{\hat M_b}= 0.15 & 0.32 & 0.77 & 1.33 & 1.29 & 0.79 & 1.68 & 0.58 \\
\frac{E_D}{\hat M_c}= 0.06,\,
\frac{E_B}{\hat M_b}= 0.02 & 0.45 & 0.82 & 0.86 & 0.87 & 0.72 & 0.99 & 0.62
\end{array}
\end{equation}
The first line is obtained using the actual values of our model parameters,
Eq.~(\ref{tableA}), while the second uses artificially inflated
constituent-quark masses: $\hat M_c= 1.76\,$GeV, $\hat M_b=5.18\,$GeV, and
$M_\rho=0$.  This comparison indicates that the scaling laws fail in our
model because $c$- and $b$-quark masses that are consistent with contemporary
estimates~\cite{pdg98} do not validate the approximations $E_H/m_H =0 =
M_\rho/m_H$ used in the derivation of these laws.

As noted in Sec.~\ref{raddecays}, in the heavy-quark limit the radiative
decays can be used to define an inertial-mass for the quarks.  We find
\begin{equation}
\begin{array}{c|ddd}
                      & \hat M_Q J^Q(0) & M_Q^{\rm in} ({\rm GeV}) 
                                & M_q^{\rm in} ({\rm GeV}) \\\hline
D^\ast \to D      & 0.56 & 2.4 & 1.5 \\
B^\ast \to B      & 0.70 & 6.7 & 1.7 \\
D_s^\ast \to D_s  & 0.85 & 1.5 & 2.3 \\
B_s^\ast \to B_s  & 1.1  & 4.2 & 2.7 
\end{array}
\end{equation}
Comparing these results with Eqs.~(\ref{tableB}) and~(\ref{usME}) it is clear
and unsurprising that $M^{\rm in}$ is not a good measure of the constituent
mass when the binding energy is large.

\section{Epilogue}
\label{sech}
We have described a direct extension of Dyson-Schwinger equation (DSE) based
phenomenology to experimentally accessible heavy-meson observables.  In doing
this we explored the fidelity of a simple approximation, Eq.~(\ref{dsehq}),
to the dressed-heavy-quark propagator.  In our framework this approximation
is a necessary precursor to effecting a heavy-quark expansion.  However, in
contrast to Refs.~\cite{mishaB,mishaPLB}, herein we elected not to consummate
that expansion and thereby achieved a {\it unified} and uniformly accurate
description of an extensive body of light- and heavy-meson observables.  In
doing so our results indicate that corrections to the heavy-quark symmetry
limit of $\lsim 30$\% are encountered in $b\to c$ transitions and that these
corrections can be as large as a factor of two in $c\to d$ transitions.

Our calculation of the semileptonic transition form factors for $B$- and
$D$-mesons on their entire kinematic domain and with the light-quark sector
well constrained is potentially useful in the experimental extraction of the
CKM matrix elements $V_{cb}$, $V_{ub}$.  That is true too of our calculation
of the leptonic decay constants; e.g., accurate knowledge of $f_B$ can assist
in the determination of $V_{td}$.  They also indicate that $D_f$-mesons do
not lie on the heavy-quark $1/\surd\hat m_Q$-trajectory.  In elucidating a
mass-formula valid for all nonsinglet pseudoscalar mesons, we demonstrated
that in the heavy-quark limit pseudoscalar meson masses grow linearly with
the mass of their heaviest constituent; i.e., $m_P \propto \hat m_Q$.  Our
calculations also provide an estimate of the total width of the
$D_{(s)}^{\ast +}$- and $D^{\ast 0}$-mesons, for which currently there are
only experimental upper-bounds.

Although this study is a significant improvement and extension of
Refs.~\cite{mishaB,mishaPLB}, more is possible.  One simple step is a wider
study of light vector meson observables so as to more tightly constrain their
properties.  Existing models for the Bethe-Salpeter kernel are applicable to
these systems and studies akin to Ref.~\cite{mr97} are underway.  A pleasing
aspect of our study, however, is the demonstration that DSE phenomenology as
applied extensively hitherto is well able to describe vector meson
polarisation observables.  A more significant extension is the development of
a kernel applicable to the study of heavy-meson masses.  That would provide
further insight into the structure of heavy-meson bound state amplitudes.
They are an integral part of our calculations but only rudimentary models are
currently available.  It would also assist in constraining DSE phenomenology
via a comparison with calculations and models of the heavy-quark potential.

\section*{Acknowledgments}
We acknowledge helpful correspondence with P. Maris and M. A. Pichowsky.  MAI
gratefully acknowledges the hospitality and support of the Physics Division
at ANL during a visit in which some of this work was conducted.  This work
was supported in part by the Russian Fund for Fundamental Research, under
contract numbers 97-01-01040 and 99-02-17731-a, and the US Department of
Energy, Nuclear Physics Division, under contract number W-31-109-ENG-38, and
benefited from the resources of the National Energy Research Scientific
Computing Center.

\appendix
\section{Collected Formulae}
The impulse approximation to the hadronic tensor describing the strong decay
of a heavy vector meson is ($\ell_{1}= \ell-p_{1}$, $q= p_1-p_2$,
$q^2=-m_\pi^2$)
\begin{eqnarray}
\label{HsHpi}
\lefteqn{M_\mu^{H^\ast H \pi}(p_1,p_2) = }\\
&& \nonumber 
\!\!\! 2 \sqrt{2} \,N_c\,{\rm tr}_D\,\int_\ell^\Lambda\,
\bar\Gamma_P(\ell;-p_2)\,S_{Q}(\ell_1) \,
i\Gamma_\mu^{H^\ast}(\ell;p_1)\,S_{u}(\ell)\,
\bar\Gamma_\pi(\ell;-q)\,S_u(\ell+q)\,,
\end{eqnarray}
and that for a light vector meson is similar:
\begin{eqnarray}
\label{VPpi}
\lefteqn{M_\mu^{VP\pi}(k_1,k_2) = }\\
&& \nonumber 2 N_c\,{\rm tr}_D\,\int_\ell^\Lambda\,
i\Gamma^V_\mu(\ell;P)\,
S_q(\ell_{++})\,\bar\Gamma_P(\ell_{0+};-k_1)\,
S_u(\ell_{+-})\,\bar\Gamma_\pi(\ell_{-0};-k_2)\,
S_u(\ell_{--})\,,
\end{eqnarray}
where $P=k_1+k_2$, $\ell_{\alpha\beta}= \ell + \case{\alpha}{2} k_1 +
\case{\beta}{2} k_2$, and $q=u,s$ for $V=\rho,K^\ast$.

The impulse approximation to the hadronic tensor describing the rare neutral
current process is ($\ell_{1,2} = \ell+p_{1,2})$
\begin{eqnarray}
\label{rareB}
\lefteqn{M_{\mu\nu}^{B\to K^\ast \gamma}(p_1,p_2) =}\\
&& \nonumber 2 N_c\,{\rm tr}_D\,\int_\ell^\Lambda\,
\bar\Gamma_\nu^{K^\ast}(\ell;-p_2)\,
S_s(\ell_2)\,
q_\rho\sigma_{\mu\rho}(1+\gamma_5)
S_b(\ell_1)\,\Gamma_{\bar B}(\ell;p_1)\,S_q(\ell)\,.
\end{eqnarray}

The heavy-quark symmetry limits of the leptonic heavy$\,\to\,$light-meson
transition form factors are Eq.~(\ref{fpmhq}) and 
\begin{eqnarray}
&&  \label{a1hq}
A_1(t) = \frac{N_c \kappa_f}{4\pi^2}
\int_0^\infty\,du\,\left(\surd u -E_H\right)\,
\varphi_H(z)\,\varphi^V_l(z)\,
\int_0^1\,d\tau\,
\frac{1}{\surd \hat M_Q} H_{A_1},\\
\nonumber \lefteqn{A_2(t)  = }\\
&&  \frac{N_c \kappa_f}{2\pi^2}
\int_0^\infty\,du\,\left(\surd u -E_H\right)\,
\varphi_H(z)\,\varphi^V_l(z)\,
\int_0^1\,d\tau\,\surd \hat M_Q\,
\left[\frac{1}{\hat M_Q} H_{A_2} + H_{A_3}\right],\\
\label{a2hq}
\nonumber \lefteqn{A_3(t)  = }\\
&&  \frac{N_c \kappa_f}{2\pi^2}
\int_0^\infty\,du\,\left(\surd u -E_H\right)\,
\varphi_H(z)\,\varphi^V_l(z)\,
\int_0^1\,d\tau\,\surd \hat M_Q\,
\left[\frac{1}{\hat M_Q} H_{A_2} - H_{A_3}\right],\\
\label{a3hq}
&&  V(t)  = \frac{N_c \kappa_f}{2\pi^2}
\int_0^\infty\,du\,\left(\surd u -E_H\right)\,
\varphi_H(z)\,\varphi^V_l(z)\,
\int_0^1\,d\tau\,\surd \hat M_Q\, H_{2}\,,
\label{vhq}
\end{eqnarray}
where: $z=u-2 E_H \surd u$, $z_1= z+2 X \tau \surd u$, $X= (\hat M_Q/2)\,(1 -
t/\hat M_Q^2)$, $z_2= (1-\tau) (z + \tau u)$, and
\begin{eqnarray}
&& H_1= \sigma_S \tilde\sigma_S 
- \tau\surd u\,\left(\sigma_S\tilde\sigma_V - \sigma_V\tilde\sigma_S\right)
+ z \sigma_V\tilde\sigma_V\,,\\
&& H_2 = \sigma_S \tilde\sigma_V
 + z_2\left(\sigma_S\tilde\sigma_V^\prime 
                - \sigma_V\tilde\sigma_S^\prime\right)
 + \tau\surd u \,\sigma_V\tilde\sigma_V\,,\\
\nonumber\lefteqn{H_{A_1} = \sigma_S \tilde\sigma_S 
+ X \left(\sigma_S \tilde\sigma_V 
        + z_2 \left[\sigma_S\tilde\sigma_V^\prime 
                - \sigma_V\tilde\sigma_S^\prime\right]\right)}\\
&& \;\;\;\;\;\;\;\;
-\tau\surd u\,\left(\sigma_S\tilde\sigma_V - \sigma_V\tilde\sigma_S\right)
+ \left( X \tau\surd u + z + z_2\right)\sigma_V\tilde\sigma_V\,,\\
&& H_{A_2} = -2 \tau\surd u\,\left(\sigma_S \tilde\sigma_V
        + \tau\surd u\,\sigma_V\tilde\sigma_V\right)\,,\\
&& H_{A_3} = H_2 
+ 2 \tau\surd u \,z_2 \sigma_V\tilde\sigma_V^\prime\,,
\end{eqnarray}
with: $\sigma=\sigma(z)$, $\tilde\sigma=(z_1)$, $\sigma^\prime=
d\sigma(z)/dz$ and $\tilde\sigma^\prime= d\sigma(z_1)/dz_1$.  

At $t=0$ the behaviour of these form factors simplifies further, as described
by Eq.~(\ref{fpt0hq}) and
\begin{eqnarray}
\nonumber \lefteqn{4 A_1(0) = A_2(0) = -A_3(0) = V(0) =}\\
&& \label{otherst0hq}
 \frac{\ln  m_H}{ m_H^{1/2}}\,
\frac{N_c \kappa_f}{2\pi^2}
\int_0^\infty\,du\,\left(1 - E_H/\surd u \right)\,
\varphi_H(z)\,\varphi^V_l(z)\,\sigma^f_S(z) \,. 
\end{eqnarray}
In Eqs.~(\ref{a1hq}) -~(\ref{otherst0hq}) all dimensioned quantities are
expressed in units of our mass-scale, $\lambda$.


%
%
\begin{table}[t]
\caption{The 16 dimension-GeV quantities used in $\chi^2$-fitting our
parameters.  The weighting error is the experimental error or 10\% of the
experimental value, if that is greater, since we expect that to be the limit
of our model's accuracy.  The values in the ``Obs.'' column are taken from
Refs.~\protect\cite{mr97,pdg98,flynn}.
\label{tableC} }
\begin{tabular}{lll|lll}
        & Obs.  & Calc. & & Obs.  & Calc. \\\hline
$f_\pi$   & 0.131 & 0.146 & $m_\pi$   & 0.138 & 0.130 \\
$f_K  $   & 0.160 & 0.178 & $m_K$     & 0.496 & 0.449 \\
$\langle \bar u u\rangle^{1/3}$ & 0.241 & 0.220 &
        $\langle \bar s s\rangle^{1/3}$ & 0.227 & 0.199\\
$\langle \bar q q\rangle_\pi^{1/3}$ & 0.245 & 0.255& 
        $\langle \bar q q\rangle_K^{1/3}$ & 0.287 & 0.296\\
$f_\rho$   & 0.216      & 0.163 & 
        $f_{K^\ast}$   & 0.244      & 0.253 \\
$\Gamma_{\rho\pi\pi}$ & 0.151     & 0.118 & 
        $\Gamma_{K^\ast (K\pi)}$  & 0.051     & 0.052 \\
$f_D$   & 0.200 $\pm$ 0.030     & 0.213 & 
        $f_{D_s}$ & 0.251 $\pm$ 0.030     & 0.234 \\
$f_B$   & 0.170 $\pm$ 0.035      & 0.182 & 
$g_{B K^\ast \gamma} \hat M_b$ & 2.03 $\pm$ 0.62 & 2.86 \\\hline
\end{tabular}
\end{table}
\begin{table}[t]
\caption{The 26 dimensionless quantities used in $\chi^2$-fitting our
parameters.  The weighting error is the experimental error or 10\% of the
experimental value, if that is greater, since we expect that to be the limit
of accuracy of our model.  The values in the ``Obs.'' column are taken from
Refs.~\protect\cite{pdg98,richman,pirad,cesr96,latt,gHsHpi}.  The light-meson
electromagnetic form factors are calculated in impulse
approximation~\protect\cite{cdrpion,brtpik,mrpion} and $\xi(w)$ is obtained
from $f_+^{B\to D}(t)$ via Eq.~(\protect\ref{fxi}).
\label{tableD} }
\begin{tabular}{lll|lll}
        & Obs.  & Calc. & & Obs.  & Calc. \\\hline
$f_+^{B\to D}(0)$ & 0.73 & 0.58  &
        $f_\pi r_\pi$ & 0.44 $\pm$ 0.004 & 0.44   \\
$F_{\pi\,(3.3\,{\rm GeV}^2)}$ & 0.097 $\pm$  0.019  & 0.077 &
        B$(B\to D^\ast)$ & 0.0453 $\pm$ 0.0032 & 0.052\\
$\rho^2$ &  1.53 $\pm$ 0.36 & 1.84 &
        $\alpha^{B\to D^\ast}$ & 1.25 $\pm$ 0.26 & 0.94 \\
$\xi(1.1)$  & 0.86 $\pm$ 0.03& 0.84 &
        $A_{\rm FB}^{B\to D^\ast}$ & 0.19 $\pm$ 0.031 & 0.24 \\
$\xi(1.2)$  & 0.75 $\pm$ 0.05& 0.72 &
        B$(B\to \pi)$ & (1.8 $\pm$ 0.6)$_{\times 10^{-4}}$  & 2.2 \\
$\xi(1.3)$  & 0.66 $\pm$ 0.06& 0.63 &
        $f^{B\to \pi}_{+\,(14.9\,{\rm GeV}^2)}$ & 0.82 $\pm$ 0.17 & 0.82 \\
$\xi(1.4) $ & 0.59 $\pm$ 0.07& 0.56 &
        $f^{B\to \pi}_{+\,(17.9\,{\rm GeV}^2)}$ & 1.19 $\pm$ 0.28 & 1.00 \\
$\xi(1.5) $ & 0.53 $\pm$ 0.08& 0.50 &
        $f^{B\to \pi}_{+\,(20.9\,{\rm GeV}^2)}$ & 1.89 $\pm$ 0.53 & 1.28 \\
B$(B\to D)$ & 0.020 $\pm$ 0.007 & 0.013 &
        B$(B\to \rho)$ & (2.5 $\pm$ 0.9)$_{\times 10^{-4}}$ & 4.8 \\
B$(D\to K^\ast)$ & 0.047 $\pm$ 0.004  & 0.049 &
        $f_+^{D\to K}(0)$ & 0.73 &  0.61 \\
$\frac{V(0)}{A_1(0)}^{D \to K^\ast}$ & 1.89 $\pm$ 0.25 & 1.74 &
        $f_+^{D\to \pi}(0)$ & 0.73 &  0.67 \\
$\frac{\Gamma_L}{\Gamma_T}^{D \to K^\ast}$ & 1.23 $\pm$ 0.13 & 1.17 &
        $g_{B^\ast B\pi}$ & 23.0 $\pm$ 5.0 & 23.2 \\
$\frac{A_2(0)}{A_1(0)}^{D \to K^\ast}$ & 0.73 $\pm$ 0.15 & 0.87 &
        $g_{D^\ast D\pi}$ & 10.0 $\pm$ 1.3 & 11.0 \\\hline
\end{tabular}
\end{table}
\begin{table}[t]
\caption{Calculated values of a range of observables not included in fitting
the model's parameters, with the light-meson electromagnetic form factors
calculated in impulse approximation~\protect\cite{cdrpion,brtpik,mrpion}.
The ``Obs.''  values are extracted from
Refs.~\protect\cite{pdg98,richman,flynn,kaondat}.
\label{tableE} }
\begin{tabular}{lll|lll}
        & Obs.  & Calc. & & Obs.  & Calc. \\\hline
$f_K r_K$       &   0.472 $\pm$ 0.038 & 0.46 &     
        $-f_K^2 r_{K^0}^2$ &  (0.19 $\pm$ 0.05)$^2$ & (0.10)$^2$   \\
$g_{\rho\pi\pi}$ & 6.05 $\pm$ 0.02 & 5.27  &
        $\Gamma_{D^{\ast 0} }$ (MeV)& $ < 2.1 $  & 0.020   \\
$g_{K^\ast K \pi^0}$ & 6.41 $\pm$ 0.06 & 5.96  & 
        $\Gamma_{D^{\ast +}}$ (keV) &  $< 131$ & 37.9 \\
$g_\rho$ & 5.03 $\pm$ 0.012 & 5.27  & 
        $\Gamma_{D_s^{\ast} D_s \gamma}$ (MeV)& $< 1.9$  & 0.001    \\
$f_{D^\ast}$ (GeV) &   & 0.290   & 
        $\Gamma_{B^{\ast +} B^+ \gamma}$ (keV)&   & 0.030    \\
$f_{D^\ast_s}$ (GeV)&   & 0.298   & 
        $\Gamma_{B^{\ast 0} B^0 \gamma}$ (keV)&  &  0.015 \\
$f_{B_s}$ (GeV) & 0.195 $\pm$ 0.035 & 0.194  & 
        $\Gamma_{B_s^{\ast} B_s \gamma}$ (keV)&  &  0.011 \\
$f_{B^\ast}$ (GeV)&   & 0.200   &
        B($D^{\ast +}\!\to D^+ \pi^0$) & 0.306 $\pm$ 0.025 & 0.316 \\
$f_{B^\ast_s}$ (GeV)&   & 0.209   & 
        B($D^{\ast +}\!\to D^0 \pi^+$) & 0.683 $\pm$ 0.014 & 0.683 \\
$f_{D_s}/f_D$ & 1.10 $\pm$ 0.06 &  1.10  &
        B($D^{\ast +}\!\to D^+ \gamma$) & 0.011~$^{+ 0.021}_{-0.007}$ & 0.001 \\
$f_{B_s}/f_B$  & 1.14 $\pm$ 0.08  & 1.07   &
        B($D^{\ast 0}\!\to D^0 \pi^0$) &  0.619 $\pm$ 0.029 & 0.826 \\
$f_{D^\ast}/f_D$ &       &  1.36  &
        B($D^{\ast 0}\!\to D^0 \gamma$) &  0.381 $\pm$ 0.029 &  0.174 \\
$f_{B^\ast}/f_B$  &       & 1.10   &
        B($B \to K^\ast \gamma$) & (5.7 $\pm$ 3.3)$_{10^{-5}}$ & 11.4 \\\hline
\end{tabular}
\end{table}
\begin{table}[t]
\caption{Calculated values of some $b\to c$ and $b\to u$ transition form
factor observables not included in fitting the model's parameters.  The
``Obs.''  values are extracted from Refs.~\protect\cite{pdg98,richman}.
\label{tableF} }
\begin{tabular}{lll|lll}
        & Obs.  & Calc. & & Obs.  & Calc. \\\hline
$A_1^{B\to D^\ast}(0)$ &   & 0.57   & 
        $A_1^{B\to D^\ast}(t_{\rm max}^{B\to D^\ast})$ &   & 0.88   \\
$A_2^{B\to D^\ast}(0)$ &   & 0.56   & 
        $A_2^{B\to D^\ast}(t_{\rm max}^{B\to D^\ast})$ &   & 1.16   \\
$V^{B\to D^\ast}(0)$ &   & 0.70   & 
        $V^{B\to D^\ast}(t_{\rm max}^{B\to D^\ast})$ &   & 1.47   \\
$R_1^{B\to D^\ast}(1)$ &   1.30 $\pm$ 0.39   & 1.32 & 
        $R_2^{B\to D^\ast}(1)$ & 0.64 $\pm$ 0.29 & 1.04   \\
$R_1^{B\to D^\ast}(w_{\rm max})$ &      & 1.23 & 
        $R_2^{B\to D^\ast}(w_{\rm max})$ &  & 0.98   \\
$\alpha^{B\to\rho}$ &   & 0.60 &
        $f_+^{B\to D}(t_{\rm max}^{B\to D})$ &   & 1.21   \\
$f_+^{B\to \pi}(0)$ & & 0.45 &
        $f_+^{B\to \pi}(t_{\rm max}^{B\to \pi})$ & & 3.73 \\
$A_1^{B\to \rho}(0)$ &   & 0.47   &
        $A_1^{B\to \rho}(t_{\rm max}^{B\to \rho})$ &   & 0.45   \\
$A_2^{B\to \rho}(0)$ &   & 0.50   & 
        $A_2^{B\to \rho}(t_{\rm max}^{B\to \rho})$ &   & 0.81   \\
$V^{B\to \rho}(0)$   &   & 0.68   & 
        $V^{B\to \rho}(t_{\rm max}^{B\to \rho})$ &   & 1.17   \\
$R_1^{B\to \rho}(1)$ &      & 1.15 & 
        $R_2^{B\to \rho}(1)$ &  & 0.80   \\
$R_1^{B\to \rho}(w_{\rm max})$ &      & 1.44 & 
        $R_2^{B\to \rho}(w_{\rm max})$ &  & 1.06   \\\hline
\end{tabular}
\end{table}

\begin{table}[t]
\caption{Calculated values of some $c\to s$ and $c\to d$ transition form
factor observables not included in fitting the model's parameters.  The
``Obs.''  values are extracted from Refs.~\protect\cite{richman,pdg98}.
\label{tableG} }
\begin{tabular}{lll|lll}
        & Obs.  & Calc. & & Obs.  & Calc. \\\hline
B($D^+\to \rho^0$) &   & 0.032   &
        $\alpha^{D \to \rho}$ &   & 1.03  \\
${\rm B}(D^0\to K^-)$   &  0.037 $\pm$ 0.002  &  0.036   &
        $\frac{{\rm B}(D\to \rho^0)}
        {{\rm B}(D\to K^\ast)}$ & 0.044 $\pm$ 0.034  & 0.065   \\
$A_1^{D\to K^\ast}(0)$ & 0.56 $\pm$ 0.04 & 0.46   & 
        $A_1^{D\to K^\ast}(t_{\rm max}^{D\to K^\ast})$ &0.66 $\pm$ 0.05 & 0.47 \\
$A_2^{D\to K^\ast}(0)$ & 0.39 $\pm$ 0.08 & 0.40   & 
      $A_2^{D\to K^\ast}(t_{\rm max}^{D\to K^\ast})$ & 0.46 $\pm$ 0.09 & 0.44 \\
$V^{D\to K^\ast}(0)$ & 1.1 $\pm$ 0.2 & 0.80   & 
        $V^{D\to K^\ast}(t_{\rm max}^{D\to K^\ast})$ & 1.4 $\pm$ 0.3 & 0.92 \\
$R_1^{D\to K^\ast}(1)$ &      & 1.72 & 
        $R_2^{D\to K^\ast}(1)$ &  & 0.83   \\
$R_1^{D\to K^\ast}(w_{\rm max})$ &      & 1.74 & 
        $R_2^{D\to K^\ast}(w_{\rm max})$ &  & 0.87   \\
$\frac{{\rm B}(D^0\to \pi)}
        {{\rm B}(D^0\to K)}$ & 0.103 $\pm$ 0.039  & 0.098   &
        $f_+^{D\to K}(t_{\rm max}^{D\to K})$ & 1.31 $\pm$ 0.04 & 1.11 \\
$\frac{f_+^{D\to \pi}(0)}{f_+^{D\to K}(0)}$ & 1.2 $\pm$ 0.3 &  1.10  &
        $f_+^{D\to \pi}(t_{\rm max}^{D\to \pi})$ &  & 2.18 \\
$A_1^{D\to \rho}(0)$ &   & 0.60   &
        $A_1^{D\to \rho}(t_{\rm max}^{D\to \rho})$  && 0.58 \\
$A_2^{D\to \rho}(0)$ &   & 0.54   &
        $A_2^{D\to \rho}(t_{\rm max}^{D\to \rho})$  && 0.64 \\
$V^{D\to \rho}(0)$   &   & 1.22   &
        $V^{D\to \rho}(t_{\rm max}^{D\to \rho})$ & & 1.51 \\
$R_1^{D\to \rho}(1)$ &      & 2.08 & 
        $R_2^{D\to \rho}(1)$ &  & 0.88   \\
$R_1^{D\to \rho}(w_{\rm max})$ &      & 2.03 & 
        $R_2^{D\to \rho}(w_{\rm max})$ &  & 0.91   \\\hline
\end{tabular}
\end{table}

\begin{thebibliography}{99}
%
\bibitem{peter} P. C. Tandy, Prog. Part. Nucl. Phys. {\bf 39}, 117 (1997).
%
\bibitem{cdranu} C. D. Roberts, ``Nonperturbative QCD with Modern Tools'',
nucl-th/9807026. 
%
\bibitem{review} C. D. Roberts and A. G. Williams,
Prog. Part. Nucl. Phys. {\bf 33}, 477 (1994).
%
\bibitem{cdranom} C. D. Roberts, R. T. Cahill and J. Praschifka,
Ann. Phys. {\bf 188}, 20 (1988).
%
\bibitem{pipi} C. D. Roberts, R. T. Cahill, M. E. Sevior and N. Iannella,
     Phys. Rev. D {\bf 49}, 125 (1994).
%
\bibitem{cdrpion} C. D. Roberts, Nucl. Phys. A {\bf 605}, 475 (1996).
%
\bibitem{brtpik} C. J. Burden, C. D. Roberts and M. J. Thomson, Phys. Lett. B
{\bf 371}, 163 (1996).
%
\bibitem{mrpion} P. Maris and C. D. Roberts, Phys. Rev. C {\bf 58}, 3659
(1998). 
%
\bibitem{photopi} M. R. Frank, K. L. Mitchell, C. D. Roberts and P. C. Tandy,
Phys. Lett. B {\bf 359}, 17 (1995); R. Alkofer and C. D. Roberts,
Phys. Lett. {\it ibid} {\bf 369}, 101 (1996).
%
\bibitem{dubravko} D. Kekez, B. Bistrovic and D. Klabucar,
Int. J. Mod. Phys. A {\bf 14}, 161 (1999).
%
\bibitem{pichowsky} M. A. Pichowsky and T.-S. H. Lee, Phys. Lett. B {\bf
379}, 1 (1996); M. A. Pichowsky and T.-S. H. Lee, Phys. Rev. D {\bf 56}, 1644
(1997).
%
\bibitem{ph98} F. T. Hawes and M. A. Pichowsky, Phys. Rev. C {\bf 59}, 1743
(1999).
%
\bibitem{mr97} P. Maris and C. D. Roberts, Phys. Rev. C {\bf 56}, 3369
(1997); P. Maris, C. D. Roberts and P. C. Tandy, Phys. Lett. B {\bf 420}, 267
(1998).
%
\bibitem{derek} D. B. Leinweber, Ann. Phys. (N.Y.) {\bf 254}, 328 (1997).
%
\bibitem{mr98} P. Maris and C. D. Roberts, ``Differences between Heavy and
Light Quarks'', in {\em Rostock 1997, Progress in heavy quark physics},
edited by M. Beyer, T. Mannel and H. Schr\"oder; nucl-th/9710062.
%
\bibitem{vary97} A. A. El-Hady, A. Datta, K. S. Gupta and J. P. Vary,
Phys. Rev. D {\bf 55}, 6780 (1997).
%
\bibitem{neubert94} M. Neubert, Phys. Rep. {\bf 245}, 259 (1994); M. Neubert,
``Heavy quark masses, mixing angles, and spin flavor symmetry'', in {\em The
Building Blocks of Creation: From Microfermis to Megaparsecs}, edited by S.
Raby and T. Walker (World Scientific, Singapore 1994), hep-ph/9404296; and
references therein.
%
\bibitem{mishaB} M. A. Ivanov, Yu. L. Kalinovsky, P. Maris and C. D. Roberts,
Phys. Rev. C~{\bf 57}, 1991 (1998).
%
\bibitem{braun} V. Braun, ``Light cone sum rules'', in {\em Rostock 1997,
Progress in heavy quark physics}, edited by M. Beyer, T. Mannel and
H. Schr\"oder, hep-ph/9801222.
%
\bibitem{pdg98} Particle Data Group (C. Caso {\it et al}.), Eur. Phys. J. C
{\bf 3}, 1 (1998).
%
\bibitem{marisPC} P. Maris, private communication.
%
\bibitem{mishaPLB} M. A. Ivanov, Yu. Kalinovsky, P. Maris and C. D. Roberts
     Phys. Lett. B {\bf 416}, 29 (1998).
%
\bibitem{richman} J. D. Richman and P. R. Burchat, Rev. Mod. Phys. {\bf 67},
893 (1995).
%
\bibitem{flynn} J. M. Flynn and C. T. Sachrajda, ``Heavy Quark Physics From
Lattice QCD'', hep-lat/9710057.
%
\bibitem{munczekburden} H. Munczek, Phys. Lett. B {\bf 175}, 215 (1986);
C. J. Burden, C. D. Roberts and A. G. Williams, {\it ibid} {\bf 285}, 347
(1992). 
%
\bibitem{brs96} A. Bender, C. D. Roberts and L. v. Smekal, Phys. Lett. B {\bf
380} (1996) 7; C. D. Roberts, in {\it Quark Confinement and the Hadron
Spectrum II}, edited by N. Brambilla and G. M. Prosperi (World Scientific,
Singapore, 1997), pp. 224-230.
%
\bibitem{mitchell} Yu. Kalinovsky, K. L. Mitchell and C. D. Roberts,
Phys. Lett. B {\bf 399}, 22 (1997).
%
\bibitem{piloop} R. Alkofer, A. Bender and C. D. Roberts,
Int. J. Mod. Phys. A {\bf 10}, 3319 (1995).
%
\bibitem{jm93} P. Jain and H. J. Munczek, Phys. Rev. D {\bf 48}, 5403
(1993). 
%
\bibitem{conrad} C. J. Burden and D. S. Liu, Phys. Rev. D {\bf 55}, 367
(1997).
%
\bibitem{ayse97} A. Bashir, A. Kizilersu and M. R. Pennington, Phys. Rev. D
{\bf 57}, 1242 (1998); and references therein.
%
\bibitem{bc80} J. S. Ball and T.-W. Chiu, Phys. Rev. D {\bf 22}, 2542 (1980).
%
\bibitem{cp92} D. C. Curtis and M. R. Pennington, Phys. Rev. D {\bf 46}, 2663
(1992). 
%
\bibitem{hawes} F. T. Hawes, C. D. Roberts and A. G. Williams, Phys. Rev. D
{\bf 49}, 4683 (1994).
%
\bibitem{buchalla} G. Buchalla, A. J. Buraz and M. E. Lautenbacher,
Rev. Mod. Phys. {\bf 68}, 1125 (1996). 
%
\bibitem{IW90} N. Isgur and M. B. Wise, Phys. Lett. B {\bf 232}, 113 (1989);
{\bf 237}, 527 (1990). 
%
\bibitem{pirad} S. R. Amendolia {\it et al}., Nucl. Phys. B {\bf 277}, 168
(1986); C. J. Bebek, {\it et al}., Phys. Rev. D {\bf 13}, 25 (1976), {\it
ibid} {\bf 17}, 1693 (1978); 
%
\bibitem{cesr96} CLEO Coll. (J. E. Duboscq {\it et al}.),
Phys. Rev. Lett. {\bf 76}, 3899 (1996).
%
\bibitem{latt} UKQCD Coll. (D. R. Burford {\it et al}.), Nucl. Phys. B
{\bf 447}, 425 (1995).
%
\bibitem{gHsHpi} V. M. Belyaev, V. M. Braun, A. Khodjamirian and R. R\"uckl,
Phys. Rev. D {\bf 51}, 6177 (1995).
%
\bibitem{simon} For example, S. Capstick and B. D. Keister, ``Baryon magnetic
moments in a relativistic quark model'', nucl-th/9611055.
%
\bibitem{privatemike} M. A. Pichowsky, private communication.
%
\bibitem{kaondat} S. R. Amendolia, {\it et al}., Phys. Lett. B {\bf 178}, 435
(1986); W. Molzon, {\it et al}., Phys. Rev. Lett. {\bf 41}, 1213 (1978).
%
\bibitem{mishaC} M. A. Ivanov, O. E. Khomutenko and T. Mitzutani,
Phys. Rev. D {\bf 46}, 3817 (1992); M. A. Ivanov and Yu. M. Valit,
Zeit. Phys. {\bf C 67}, 633 (1995); M. A. Ivanov and Yu. M. Valit,
``Heavy-to-light form factors in the quark model with heavy
infrapropagators'', hep-ph/9606404, Few-Body Syst. (to be published).
%
\bibitem{simula} N. B. Demchuk, I. L. Grach, I. M. Narodetski and S. Simula,
Phys. Atom. Nucl. {\bf 59}, 2152 (1996).
%
\bibitem{argus93} ARGUS Collaboration, Z. Phys. C {\bf 57}, 249 (1993).
%
\end{thebibliography}
\end{document}